\documentclass[bibyear]{aa}
\usepackage{graphics}
\usepackage{graphicx}
\usepackage[varg]{txfonts}
\usepackage{psfig}
\usepackage{float}
\usepackage{psfrag}
\usepackage{amsmath}
\usepackage{amssymb}
\usepackage{ulem}
\usepackage{threeparttable}
\usepackage{array}
\usepackage{multirow}
\usepackage{txfonts}
\usepackage{enumerate}
\usepackage{xcolor}
\usepackage[T1]{fontenc}
\usepackage{natbib,twoopt}
\definecolor{darkgreen}{rgb}{0,0.5,0}
\usepackage[breaklinks=true,colorlinks=true,linkcolor=blue,citecolor=darkgreen,linkbordercolor=blue,citebordercolor=magenta,urlbordercolor=green]{hyperref}
\bibpunct{(}{)}{;}{a}{}{,}

\newcommand{\plk}{{\it Planck~}}
\newcommand{\gas}{{\mathrm{gas}}}
\newcommand{\mKs}{{\mu\mathrm{K}^2}}
\defcitealias{R12}{R12}
\defcitealias{Ar10}{A10}

\begin{document}

\title{Constraining the intracluster pressure profile\\from the thermal SZ power spectrum}
\author{M.~E.~Ramos-Ceja\thanks{Member of the International Max Planck Research School (IMPRS) for Astronomy and Astrophysics at the Universities of Bonn and Cologne.} \and K.~Basu \and F.~Pacaud \and F.~Bertoldi}
\institute{
     	Argelander-Institut f\"ur Astronomie, 
     	University of Bonn, 
     	Auf dem H\"ugel 71, 
     	53121 Bonn, 
     	Germany\\
	\email{miriam@astro.uni-bonn.de;~kbasu@astro.uni-bonn.de}
}

\abstract{
The angular power spectrum of the thermal Sunyaev-Zel'dovich (tSZ) effect is highly sensitive to cosmological parameters such as $\sigma_8$ and $\Omega_\textrm{m}$, but its use as a precision cosmological probe is hindered by the astrophysical uncertainties in modeling the gas pressure profile in galaxy groups and clusters. In this paper we assume that the relevant cosmological parameters are accurately known and explore the ability of current and future tSZ power spectrum measurements to constrain the intracluster gas pressure or the evolution of the gas mass fraction, $f_{\mathrm{gas}}$. We use the CMB bandpower measurements from the South Pole Telescope and a Bayesian Markov Chain Monte Carlo (MCMC) method to quantify deviations from the standard, universal gas pressure model. We explore analytical model extensions that bring the predictions for the tSZ power into agreement with experimental data. We find that a steeper pressure profile in the cluster outskirts or an evolving $f_{\mathrm{gas}}$ have mild-to-severe conflicts with experimental data or simulations. Varying more than one parameter in the pressure model leads to strong degeneracies that cannot be broken with current observational constraints. We use simulated bandpowers from future tSZ survey experiments, in particular a possible 2000 deg$^2$ CCAT survey, to show that future observations can provide almost an order of magnitude better precision on the same model parameters. This will allow us to break the current parameter degeneracies and place simultaneous constraints on the gas pressure profile and its redshift evolution, for example.
}


\date{Received ... ; accepted ...}

\titlerunning{ICM constraints from the tSZ power spectrum}
\authorrunning{M. E. Ramos-Ceja et al.}
\maketitle

%%%%%%%%%%%%%%%%%%%%%%%%%%%%%%%%%%%%%%%%%%%%%%%%%%%%%%%%%%%%%%%%%%%%%%%%%%%%%%%%%%%%%%%%%%%%%%%%%%%%%%%%%%%%%%%%%%%%%%%%%%%%%%%%%%%%%%%%%%%%%%%

\section{Introduction}

The scattering imprint of the cosmic microwave background (CMB) radiation from the hot, thermalized electrons in the intracluster medium (ICM) is known as the Sunyaev-Zel'dovich (SZ) effect (\citealt{SZ72}, \citeyear{SZ80}), which is playing an increasingly important role in the cosmological and astrophysical research using galaxy clusters. The SZ effect is generally divided into two distinct processes: the kinetic SZ (kSZ) effect describes the anisotropic scattering due to the cluster bulk motion, while the thermal SZ (tSZ) effect describes the inverse Compton scattering of the CMB photons by the thermal distribution of hot electrons in the ICM, which is proportional to the line-of-sight integral of the electron pressure. For the tSZ effect, the energy gain by the CMB photons gives rise to a specific spectral dependence of the tSZ signal, such that below roughly 217 GHz, clusters appear as a decrement and above as an increment in the CMB surface brightness. Because the tSZ surface brightness is independent of the redshift of the scattering source, it provides a powerful means to study the structure and dynamics of the hot intracluster gas throughout cosmic history. 

Apart from observing individual clusters, the tSZ effect can also be detected in a statistical sense through the excess power over the primordial CMB anisotropies, coming from all the resolved and unresolved galaxy groups and clusters in a CMB map. Unlike the optical and X-ray observables, the redshift independence of the tSZ signal makes this ``confusion noise'' a significant source of temperature anisotropies at millimeter/submillimeter wavelengths in the arcmin scale regime, where the primordial CMB anisotropies are damped exponentially. Similar to the cluster number counts, the tSZ anisotropy signal is sensitive to the same set of cosmological parameters because its contribution comes primarily from the hot ($\gtrsim 1$ keV), ionized ICM bound to groups and clusters \citep[e.g.,][]{Her06}. The amplitude of the tSZ power spectrum depends roughly on the eighth power of $\sigma_8$, the rms amplitude of the matter density fluctuations and on the third power of $\Omega_{\textrm{m}}$ \citep{KS02,Tr11}. However, the tSZ power receives significant contribution from the low mass, high redshifts objects. This seriously hinders its use as a precision cosmological probe, since the thermodynamic properties of these systems are not well constrained from direct observations. Thus, the astrophysical uncertainties in modeling the tSZ power spectrum are too large to place significant constraints on cosmological parameters. Consequently, attention has moved to measuring the higher order statistics of the correlation function, such as the tSZ bispectrum, which arises mostly from massive systems at intermediate redshifts and is therefore less prone to astrophysical systematics \citep[e.g.,][]{RMS03,Bh12,W12}.

As the measurement precision of the cosmological parameters improves from other methods, the use of the tSZ power spectrum in cosmology can be reversed. The tSZ power can then be used as a probe for measuring the distribution and evolution of the intracluster gas, down to low cluster masses and up to high redshifts, where direct observations are difficult. Similar arguments have been presented by several authors \citep[e.g.,][]{Sh10,Ba12a}, although a quantitative comparison between the results of analytical cluster pressure models and the observations of the tSZ power spectrum has been lacking. It is also of great interest to know how the future ground-based SZ surveys may constrain the intracluster gas models because their resolutions are better suited to constraining the shape of the tSZ power spectrum.

The amplitude of the tSZ power spectrum was predicted analytically by \citet{KS02}, with an expected value of $8-10~\mKs$ around $\ell = 3000$. Later semi-analytic modeling predicted similar values \citep{Se10}, but experimental results have confirmed these early predictions to be too high. The first conclusive measurement of the combined tSZ$+$kSZ power spectrum came from the South Pole Telescope \citep[SPT;][]{Lu10}. Successive data releases from the SPT and ACT (Atacama Cosmology Telescope) have provided increasingly sensitive and consistent measurements of the tSZ power on arcminute scales ($\ell \sim \mathrm{few}\times 1000$) where the contributions from galaxy groups and clusters are expected to peak \citep{Sh11,Du11,R12,Si13,Ge14}. Recent results from the \plk spacecraft are also consistent with the SPT value within $2\sigma$ \citep{Pl13b}, although the \plk resolution cannot resolve the position of the peak of the tSZ power, and is more sensitive on roughly degree angular scales. At these low multipoles, the two-halo correlation term might be important, or the contribution from the warm-hot intergalactic medium (WHIM) might dominate \citep[e.g.,][]{SV13}.

In this work we adopt the SPT measurement of CMB bandpowers from \citet{R12}, which constrained the peak amplitude of the tSZ power spectrum at 150 GHz at $3.65\pm0.69~\mKs$. This value is less than half of what was predicted from early semi-analytic cluster models. The source of this discrepancy has been investigated in several works, following analytic and semi-analytic modeling \citep{Sh10,Tr11}, as well as full hydrodynamical simulations \citep{Ba12a,McC14}. These authors identify several physical processes that can produce a lower amplitude of the tSZ power, namely the turbulent bulk motions of the intracluster gas, feedback from supernovae and AGN -- plus the redshift evolution of these quantities -- that cause cluster properties to deviate from a simple self-similar scaling. The uncertainties in the implementation of various astrophysical processes in these semi-analytical or numerical models remain sufficiently high \citep[$\sim 30\%$, e.g.,][]{Sh10}, such that cosmological constraints using template models for the tSZ power spectrum are generally not the most competitive.

We aim to make a detailed comparison between analytical models for the intracluster pressure and the latest tSZ power spectrum data, such that errors on the model parameters can be derived directly from observations. This contrasts with earlier analytic or semi-analytic works that were compared against simulation predictions. We set up a Markov Chain Monte Carlo (MCMC) based method to explore the range of possible values in the selected pressure model from a set of CMB bandpower measurements, obtaining the full covariance between these parameters. This also allows us to use simulated bandpowers from the future CMB/SZ experiments (e.g., CCAT and SPT-3G) to predict their ability to break the parameter degeneracies and constrain cluster physics.

This paper is organized as follows. In Section~\ref{sec:method} we describe the ``halo model'' for computing the tSZ power spectrum, followed-by its measurement technique, and outline our procedure to constrain cluster model parameters from the SPT data. Section~\ref{sec:pressure_profile} summarizes the current knowledge on the cluster pressure structure which will provide the baseline of our work. Section~\ref{sec:results} presents our attempts to reconcile the tSZ power spectrum model predictions and available measurements, using altered pressure models. In Section~\ref{sec:discussion} we extend our analysis to future SZ cluster survey experiments, and discuss the impact of cosmological parameter uncertainties on the results. We summarize our work in Section~\ref{sec:summary} and present conclusions. Throughout this work we assume a $\Lambda$CDM cosmology with $h=0.71$, $\Omega_\textrm{m}=0.264$, $\Omega_\textrm{b}=0.044$, $\Omega_{\Lambda}=0.736$, $n_{s}=0.96$, and $\sigma_{8}=0.81$ \citep{Ko11}.

%%%%%%%%%%%%%%%%%%%%%%%%%%%%%%%%%%%%%%%%%%%%%%%%%%%%%%%%%%%%%%%%%%%%%%%%%%%%%%%%%%%%%%%%%%%%%%%%%%%%%%%%%%%%%%%%%%%%%%%%%%%%%%%%%%%%%%%%%%%%%%

\section{Method}
\label{sec:method}

In this Section we describe the halo formalism for computing the tSZ power spectrum, and the observation and modeling of the microwave sky that we use in a Bayesian MCMC formalism to constrain cluster pressure model parameters.

\subsection{Analytical estimate of the tSZ power spectrum}
\label{subsec:SZcompute}

\subsubsection{From the halo model to the power spectrum}

The tSZ power spectrum consists of one-halo and two-halo contributions. The one-halo term results from the Comptonization profile of individual halos in a Poisson distributed population, while the two-halo term accounts for the two-point correlation function between individual halos. For intermediate to small angular scales ($\ell \gtrsim 1000$), which correspond to the angular size of individual galaxy clusters ($\theta\lesssim10^{\prime}$), \citet{KK99} showed that the one-halo Poisson term is by far the dominant contribution. Following their prescription, the analytical expression of the tSZ power reduces to the formula
\begin{equation}
 C_{\ell}=f^{2}_{\nu}(x)\int_{0}^{z_{\max}} dz \frac{dV}{dz} \int_{M_{\min}}^{M_{\max}} dM\frac{dn(M,z)}{dM}|\tilde y_{\ell}(M,z)|^{2}. 
\label{szpw_poisson}
\end{equation}
Here $dV/dz$ is the co-moving volume of the Universe per unit redshift $z$, $\tilde y_{\ell}$ is the spherical harmonics decomposition of the sky-projected Compton $y-$parameter, $dn(M,z)/dM$ is the dark matter halo mass function, $f_{\nu}(x)$ is the spectral function of the tSZ effect given by
\begin{equation}
f_{\nu}(x)=\bigg(x\frac{e^{x}+1}{e^{x}-1}-4\bigg)[1+\delta_{\textrm{SZ}}(x)],
\label{spectral_function}
\end{equation}
where $x\equiv 2\pi\hbar\nu /k_{\textrm{B}}T_{\textrm{CMB}}$, and $\delta_{\textrm{SZ}}(x)$ is the relativistic correction to the frequency dependence. The reduced Planck constant is $\hbar$, $k_{\textrm{B}}$ is the Boltzmann constant, and $T_{\textrm{CMB}}$ is the CMB temperature. We do not include relativistic corrections to $f_{\nu}(x)$ as they have a negligible effect at the temperatures of groups and low-mass clusters which dominate the tSZ power spectrum.

The integral in Eq.~\ref{szpw_poisson} is insensitive to $z\gtrsim 4$ due to the absence of sufficiently massive halos. In our calculations, we thus set the upper redshift boundary to $z_{\max}=6$. Similarly, to cover a maximum critical mass range for galaxy groups and clusters, we set $M_{\min}=10^{12}~h^{-1}$~M$_{\odot }$ and $M_{\max}=10^{16}~h^{-1}$~M$_{\odot }$. We use the mass function obtained by \citet{Ti08}. 

From \citet{KS02}, the spherical harmonics contribution of a given Compton $y-$parameter profile on angular scale $\ell$ is given by
\begin{equation}
\tilde y_{\ell}(M,z)=\frac{4\pi r_{500}}{l^{2}_{c}} \int dr^{\prime }r^{\prime~ 2}y_{\textrm{3D}}(M,z,r^{\prime })\frac{\sin(\ell r^{\prime }/l_{c})}{\ell r^{\prime }/l_{c}},
\end{equation}
where $r^{\prime }\equiv r/r_{500}$ is a scaled, non-dimensional radius, $l_{c}\equiv D_{\textrm{A}}/r_{500}$ is the corresponding angular wavenumber, and $y_{\textrm{3D}}$ is the 3D radial profile of the Compton $y-$parameter. This last parameter is given by a thermal gas pressure profile, $P_{\textrm{gas}}$, through
\begin{eqnarray}
y_{\textrm{3D}}(M,z,r^{\prime}) &\equiv& \frac{\sigma_{T}}{m_{e}c^{2}}P_{e}(M,z,r^{\prime})=\frac{\sigma_{\textrm{T}}}{m_{e}c^{2}}\bigg(\frac{2+2X}{3+5X}\bigg)P_{\textrm{gas}}(M,z,r^{\prime})\nonumber \\
 &=& 1.04\times 10^{4}~\textrm{Mpc}^{-1}\bigg[\frac{P_{\textrm{gas}}(M,z,r^{\prime})}{50~\textrm{eV cm}^{-3}}\bigg],
\label{y_3D}
\end{eqnarray}
where $P_{e}$ is an electron-pressure profile, $X=0.76$ is the Hydrogen mass fraction, $\sigma_{\textrm{T}}$ is the Thomson cross-section, $m_{e}$ is the electron mass, and $c$ is the speed of light.

\subsubsection{Effect of the intrinsic pressure scatter}
\label{subsec:intrinsic_scatter}

Despite the tight correlation between the tSZ signal and cluster mass, several works have shown that the dispersion of individual cluster pressure profiles $P_e(r^{\prime})$ at a given mass is far from negligible ($\sim$30\% according to \citealt{Pl13} and \citealt{Sa13}). So far the contribution of this scatter has not been considered in analytical treatments of the tSZ power spectrum. Modeling this effect would indeed require a detailed knowledge of the diversity of cluster Comptonization morphologies. Observationally this is not yet well-constrained as tSZ experiments still aim at improving our estimate of the average cluster pressure profiles.

In our modeling of the tSZ power, we try to capture the bulk contribution of the intrinsic scatter. To do so, we make the assumption that the dispersion in the pressure structure can be encapsulated in a simple scatter on the normalization of the pressure profile, leaving the shape unchanged. In this case, marginalizing over the distribution of profile normalization results in a straightforward scaling of the tSZ power spectrum amplitude as follows:
\begin{equation}
 C_{\ell}^{s}=(1+\sigma^2_s)C_{\ell},
 \label{eq:scatter}
\end{equation}
where $\sigma_s$ is the intrinsic scatter on the $P_e(r^{\prime})$ normalization. With this approximation, the 30\% intrinsic scatter on the pressure amplitude increases the tSZ power amplitude by roughly 10\%. We include this additional contribution in all subsequent results. More details on the adopted pressure profile and the measurement of intrinsic scatter are given in Section \ref{sec:pressure_profile}.

\subsection{Microwave sky model}

Our analysis relies on the microwave extragalactic power spectra published by \citet[hereafter R12]{R12} for three different frequency bands. Those observations were extracted from $800~$deg$^{2}$ maps obtained within the SPT survey and cover angular scales of $2000<\ell<10000$.

Such observations are a combination of signals from primary CMB anisotropy, foregrounds, and secondary SZ anisotropies. The power spectrum from each of these components has different frequency dependence, so detailed multifrequency observations can in principle distinguish their relative contributions in the maps \citep[see e.g.,][]{PlanckXII}. Unfortunately, three frequencies are not sufficient to perform this kind of analysis. Instead, we have to rely on a model for the microwave sky power, calibrated using external information wherever possible. The problem then reduces to a decomposition of the observed signal into a set of templates, for which mostly the normalization has to be quantified.

In this purpose, we use the same model as in \citetalias{R12} where the total microwave sky power, $D_{\ell}^{\textrm{mod}}$, breaks down into the following components:
\begin{equation}
 D_{\ell}^{\textrm{mod}}=D_{\ell}^{\textrm{CMB}}+D_{\ell}^{\textrm{tSZ}}+D_{\ell}^{\textrm{kSZ}}+D_{\ell}^{\textrm{P}}+D_{\ell}^{\textrm{C}}+D_{\ell}^{\textrm{R}}+D_{\ell}^{\textrm{cir}}.
\label{eq:linearcombi}
\end{equation}
Here, $D_{\ell}^{\textrm{CMB}}$ is the lensed primary CMB anisotropy power at multipole $\ell$. On scales $\ell>3000$, $D_{\ell}^{\textrm{CMB}}$ is strongly damped, and other components start to dominate. The population of dusty star-forming galaxies (DSFGs) have a significant microwave emission (specially at high frequencies), which contributes with Poisson ($D_{\ell}^{\textrm{P}}$) and clustered ($D_{\ell}^{\textrm{C}}$) power components. Likewise, mainly at low frequencies, the population of radio sources contributes prominently with a Poisson term, $D_{\ell}^{\textrm{R}}$. A small contribution from the Galactic cirrus emission, $D_{\ell}^{\textrm{cir}}$, is also taken into account. Finally, the power of the tSZ and kSZ signals are given by $D_{\ell}^{\textrm{tSZ}}$ and $D_{\ell}^{\textrm{kSZ}}$, respectively.

Our work can therefore be considered as an astrophysical extension of the analysis presented by \citetalias{R12}, where we allow for more freedom in the tSZ power spectrum by tieing it to a range of empirical models of the ICM. The other components of our baseline model are treated as nuisance parameters and described in detail in Appendix~\ref{app:A}.

\subsection{Parameter estimation}

The SPT measurements as described in \citetalias{R12} comprise three auto-spectra (95$\times$95, 150$\times$150,~220$\times$220~GHz), and three cross-spectra (95$\times$150,~95$\times$220, 150$\times$220~GHz), in $15$ spectral bands $b_{\ell}$, each covering a narrow range in $\ell$. In our work, we seek to match the parameters of intracluster pressure models with those observations. This is achieved by minimization of the $\chi^{2}$ statistic,
\begin{equation}
 \chi^{2}=\sum_{b_{\ell}}\sum_{\nu_{1},\nu_{2}}(D_{b_{\ell},\nu_{1},\nu_{2}}^{\textrm{obs}}-D_{b_{\ell},\nu_{1},\nu_{2}}^{\textrm{mod}})
 ~\mathbb{N}^{-1}_{b_{\ell},\nu_{1},\nu_{2}}(D_{b_{\ell},\nu_{1},\nu_{2}}^{\textrm{obs}}-D_{b_{\ell},\nu_{1},\nu_{2}}^{\textrm{mod}})^{^T},
\end{equation}
where $D_{b_{\ell},\nu_{1},\nu_{2}}^{\textrm{obs}}$ and $D_{b_{\ell},\nu_{1},\nu_{2}}^{\textrm{mod}}$ are respectively the observed and modeled powers in band $b_{\ell}$, and $\mathbb{N}^{-1}_{b_{\ell},\nu_{1},\nu_{2}}$ is the bandpower noise covariance matrix (obtained from \citetalias{R12}), for the cross-spectra at frequencies $\nu_{1}$ and $\nu_{2}$.

The modeled bandpowers are estimated from the full-resolution power spectra of Eq.~\ref{eq:linearcombi} and the band window functions $w_{\ell,b_{\ell}}$ (also obtained from \citetalias{R12}) as:
\begin{equation}
  D_{b_{\ell}}^{\textrm{mod}} = \sum_{\ell} w_{\ell,b_{\ell}}\,\times\,D_{\ell}^\textrm{mod}.
\end{equation}
The best fit parameters and their errors are obtained by sampling the likelihood function over the whole parameter space using a MCMC Metropolis algorithm.

\begin{figure}
\begin{center}
\includegraphics[viewport= 40 10 365 335,clip,width=\columnwidth]{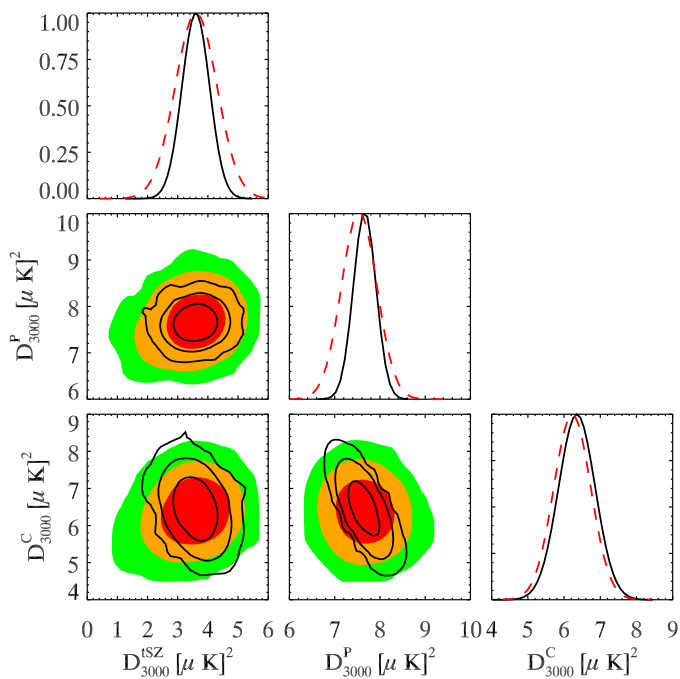}
\caption{2D likelihoods for the power spectra amplitudes at $\ell=3000$ using our MCMC algorithm, compared to the results from \citetalias{R12}. The plot shows the tSZ power spectrum ($D_{3000}^{\textrm{tSZ}}$) using the template of \citet{Sh10}, and the two CIB components, its Poisson contribution ($D_{3000}^{\textrm{P}}$) and the clustered component ($D_{3000}^{\textrm{C}}$). The filled colored contours show the 1, 2, and 3$\sigma$ constraints from the SPT analysis. The black solid contours show the constraints from our MCMC sampling, where the other foregrounds are held fixed together with the cosmological parameters.}
\label{fig:2d_SPTourcomparison}
\end{center}
\end{figure}

In order to validate our modeling of the SPT data and our fitting procedure, we first replaced our analytical tSZ model with the template provided by \citet{Sh10} - as done in \citetalias{R12} - and jointly fitted the amplitudes of the tSZ template ($D_{3000}^{\textrm{tSZ}}$), Poisson ($D_{3000}^{\textrm{P}}$) and clustered ($D_{3000}^{\textrm{C}}$) CIB components, fixing all other parameters to the \citetalias{R12} values. The results of this three-parameter samplings are displayed in Table~\ref{table_comparison} and Fig.~\ref{fig:2d_SPTourcomparison}, and are in agreement with \citetalias{R12}. As expected, the errors in our result are smaller, since we held fixed the cosmological parameters and several of the foreground components. However, the difference is not large, reflecting the fact that the three fitted components are the leading contributors to the microwave background anisotropies on the considered wavelengths and angular scales.

In the following, we always fit the amplitudes of the Poisson and clustered CIB components together with our cluster pressure model parameters. Given the small impact they have on the final measurements, the additional components (lensed CMB, kSZ effect, radio and Galactic cirrus) are kept fixed for simplicity, and so are the cosmological parameters. The cosmological constraint is discussed in more detail in Section \ref{sec:cosmology}.

%%%%%%%%%%%%%%%%%%%%%%%%%%%%%%%%%%%%%%%%%%%%%%%%%%%%%%%%%%%%%%%%%%%%%%%%%%%%%%%%%%%%%%%%%%%%%%%%%%%%%%%%%%%%%%%%%%%%%%%%%%%%%%%%%%%%%%%%%%%%%

\section{Pressure model}
\label{sec:pressure_profile}

In this Section we introduce briefly the self-similar theory that describes the properties of galaxy groups and clusters. We also explain in detail the latest measurements of the pressure profile in galaxy clusters, and show the inconsistency between the SPT measurements and the theoretical predictions for the tSZ bandpowers based on such pressure profiles.

\subsection{Self-similar models}
\label{subsec:ssmodel}

In a hierarchical structure formation scenario on a CDM cosmology, groups and clusters are the end products of gravitational collapse of a small population of highly over-dense regions in the early universe. The term ``self-similar'' points to the scale-free nature of gravitational collapse in such a universe \citep{Ka86}. This implies that if clusters were strictly self-similar, we would expect the same evolution of their global properties on any scale and time. In the context of self-similar evolution, the redshift dependence of cluster observables can be expressed as a combination of different powers of $E(z)$ and $\Delta(z)$, where $E(z) = \sqrt{\Omega_\textrm{m} (1+z)^3 + \Omega_{\Lambda}}$ is the Hubble ratio in flat $\Lambda$CDM universe, and $\Delta(z)$ is the density ratio between the mean (or critical) density of the universe at redshift $z$ and that inside a fiducial radius of the cluster.

Furthermore, a self-similar formation model implies that gravitational collapse is the only source of energy input into the ICM. Since we assume cluster formation process is governed by gravity alone, we can derive simple scaling relations for the global observables properties as a function of cluster mass. These relations have been extensively studied in the literature \citep[see][and references therein]{Boe12}. Particularly in the nearby Universe, they have been studied and determined with high precision. In the following subsection we describe the current state of knowledge on the self-similar redshift evolution and mass scaling of the ICM pressure profile.

\begin{table}
\caption[]{Comparison between the tSZ and CIB power amplitudes for our MCMC modeling and \citetalias{R12} results.}
\centering
{\small
\begin{tabular}{c c c c}
\hline
Amplitude & SPT & This work & CCAT\\
\hline
$D_{3000}^{\textrm{tSZ}}$ & $3.65\pm0.69$ & $3.61\pm0.46$ & $3.65\pm0.09$ \\
$D_{3000}^{\textrm{P}}$ & $7.54\pm0.38$ & $7.66\pm0.25$ & $7.54\pm0.02$ \\
$D_{3000}^{\textrm{C}}$ & $6.25\pm0.52$ & $6.35\pm0.51$ & $6.24\pm0.05$ \\
\hline
\end{tabular}}
\label{table_comparison}
\end{table}

\subsection{The ``universal'' pressure profile}

One of the most successful application of the self-similar model is the dark matter halo mass profile measured from N-body simulations by \citet*{Nav95}, known as the NFW profile. Following this model a more generalized version was proposed by \citet{Na07} to describe the gas distribution in galaxy clusters, which contains additional shape parameters besides the normalization and the scale radius (generalized NFW, or GNFW profile). \citet[][hereafter A10]{Ar10} measured these parameters for the GNFW model, as well as the mass scaling of the overall normalization, combining X-ray data and numerical simulations. This was the first demonstration of a scale-free, universal shape of the cluster pressure profile with a mass scaling very close to self-similar. The parametrization of the GNFW pressure model found by \citetalias{Ar10} is now commonly known as the ``universal'' pressure profile, and forms the basis of our analytical modeling.

\citetalias{Ar10} measured the GNFW profile parameters from a sample of 33 local clusters ($z<0.2$), selected from the REFLEX catalogue and observed with {\it XMM-Newton}. The sample covers a mass range $7\times 10^{13}\lesssim M_{500}h/$M$_{\odot}\lesssim 6\times 10^{14}$, where $M_{500}$ is the mass enclosed within $r_{500}$, in which the mean density is $500$ times the critical density of the Universe. The pressure profile is given by
\begin{equation}
P_{\textrm{e}}(r^{\prime})=1.65\bigg(\frac{h}{0.7}\bigg)^{2}\textrm{eV cm}^{-3}~E^{\frac{8}{3}}(z)\bigg[\frac{M_{500}}{3\times10^{14}(0.7/h)~\textrm{M}_{\odot }}\bigg]^{\frac{2}{3}+\alpha_{p}}p(r^{\prime}).
\label{A_pressure}
\end{equation}
The function $p(r^{\prime})$ describes the scale-invariant shape of the pressure profile, 
\begin{equation}
p(r^{\prime})\equiv \frac{P_{0}(0.7/h)^\frac{3}{2}}{(c_{500}r^{\prime})^{\gamma}[1+(c_{500}r^{\prime})^{\alpha}]^{(\beta-\gamma)/\alpha}},
\label{gnfw_profile}
\end{equation}
where $P_{0}=8.403$, $c_{500}=1.177$ is a gas concentration parameter, the parameter $\gamma=0.3081$ represents the central slope ($r^{\prime}<<1$), $\alpha=1.051$ is the intermediate slope ($r^{\prime}\sim 1$), and $\beta=5.4905$ is the outer slope ($r^{\prime}>>1$) of the pressure profile. The \citetalias{Ar10} profile was constrained from X-ray observations out to radii $r\sim r_{500}$. Beyond this radius, an extrapolation was made to fit the results from numerical simulations of clusters \citep{Bo04,Na07,PV08}. A small deviation from the self-similar scaling with cluster mass is given by $\alpha_{p}=0.12$. We incorporate this additional mass dependence in all our calculations, which has a small effect on the amplitude of the tSZ power ($\sim 9\%$ at $\ell=3000$). We ignore the extra smaller shape-dependent term of the mass scaling, described by the $\alpha^{\prime}_p(x)$ term \citepalias[see Eq. 10 in][]{Ar10}, because it has a negligible contribution ($\sim 2\%$ at $\ell=3000$). Finally, the derived average \citetalias{Ar10} pressure profile has a dispersion around it, which is less than $30\%$ beyond the core ($r>0.2r_{500}$), and increases towards the center. This deviation around the mean is mainly due to the dynamical state of the clusters. Following \ref{subsec:intrinsic_scatter} we assume this scatter only affects the pressure shape normalization ($P_0$), and incorporate its contribution into our power spectrum calculations accordingly.

The nearly self-similar mass scaling of the universal pressure profile has been verified down to the low mass end (galaxy groups) in the local universe. \citet{Su11} extended the measurements of \citetalias{Ar10} to lower masses, from a study of $43$ galaxy groups at $z=0.01-0.12$, within a mass range of roughly $M_{500}=10^{13}-10^{14}~h$~M$_{\odot }$. All the ICM properties of these groups were derived at least out to $r_{2500}$ from observations made with {\it Chandra}, and $23$ galaxy groups have in addition masses measured up to $\sim r_{500}$. 
As with the original data set used by \citetalias{Ar10}, the X-ray comparison by \citet{Su11} does not reveal the state of the gas pressure in groups at higher redshift, or at radii beyond $r_{500}$. Nevertheless, this rules out the possibility of a highly non-self-similar mass scaling for the ICM pressure, at least in the low redshifts, in a mass range spanning nearly two decades. Recent results by McDonald et al. (2014), using X-ray follow-up observation of SZ selected clusters, also confirm the validity of the universal pressure profile in a wide redshift range, down to a mass limit $M_{500} \sim 3\times 10^{14}$~M$_{\odot }$. These are strong constraints while finding modifications for the universal pressure model to match the tSZ power spectrum observation.

Direct tSZ observations of individual clusters have now verified the validity of the universal pressure model at the high mass end, even though measurement errors remain high. In contrast to the X-ray measurement of the GNFW profile, tSZ observations are more sensitive to the cluster outskirts ($r > r_{500}$).
\citet{Pl13} measured the pressure profile of a sample of high mass and low redshift galaxy clusters. Their mean profile shows a slightly lower pressure in the inner parts when compared with the \citetalias{Ar10} profile, although there are strong degeneracies between the derived model parameters. It is possible that the lower pressure in the core region detected by {\it Planck} is a consequence of detecting more morphologically disturbed clusters than other samples. Outside the core region the {\it Planck}-derived pressure profile show good agreement with the extrapolated \citetalias{Ar10} pressure model, out to a radius $\sim 3r_{500}$. It should be noted that the poor angular resolution of {\it Planck} restricts its sensitivity for the pressure profile shape measurement mostly to a handful of nearby, massive clusters detected with high S/N.

A higher-resolution tSZ observation of individual cluster pressure profiles became available from the {\it Bolocam} experiment \citep{Sa13}, which has 1 arcminute resolution at 150 GHz. The {\it Bolocam} team observed $45$ massive galaxy clusters, with a median mass of $M_{500}=9\times 10^{14}$~M$_{\odot}$, and spanning a large range in redshift: $0.15<z<0.89$. They fitted a GNFW profile between $0.07r_{500}$ and $3.5r_{500}$. Despite the strong covariance between the model parameters, the overall shape is fairly well constrained. The mean profile shows good agreement with the \citetalias{Ar10} model, although there is an indication of a shallower pressure outer slope ($r \gtrsim r_{500}$). Furthermore, both {\it Planck} and {\it Bolocam} teams have measured the intrinsic scatter for the pressure profile in their samples, finding it to be roughly 30\% as was also noted by \citetalias{Ar10}. In the absence of more accurate measurements, we use $\sigma_s=0.3$ (see Eq. \ref{eq:scatter}) as a fiducial value in the rest of the paper.

An important point to note here is that neither the {\it Planck} not the {\it Bolocam} analysis is based on a representative sample of galaxy clusters. Therefore, although these two data sets serve to constrain the alterations to the \citetalias{Ar10} pressure model in Section \ref{sec:results}, none can yet be considered compelling.

\begin{figure}
\begin{center}
\includegraphics[width=\columnwidth]{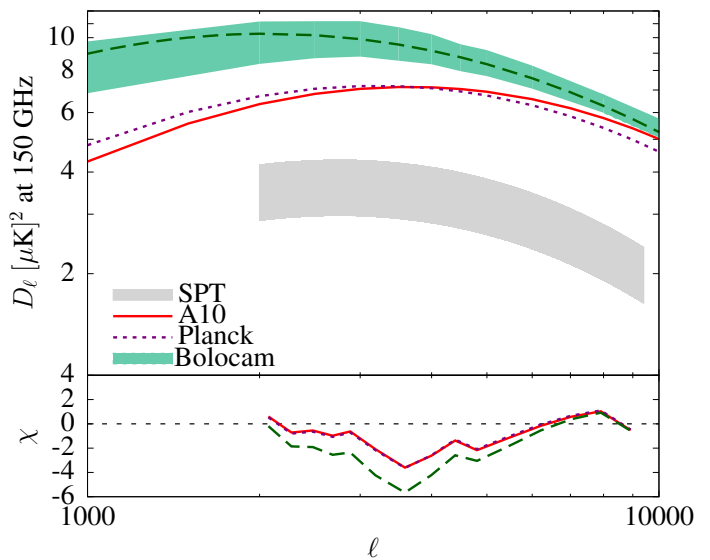}
\caption{Discrepancy between the semi-analytic model predictions for the tSZ power and the SPT measurements. The red solid line is the tSZ power spectrum given by the \citetalias{Ar10} pressure profile (Eq.~\ref{A_pressure}). The shaded gray region represent the 1$\sigma$ constraints from the SPT, which is restricted by the shape of the Shaw model (see text for details). The purple dot-dashed line gives the result for the mean GNFW profile as measured by \plk \citep{Pl13}. The green band marks the corresponding {\it Bolocam} measurement \citep{Sa13} with the 68\% confidence interval on their model fit parameters. The bottom panel shows the value of $\chi$ at each point for these models with respect to the SPT bandpower measurements, in units of the measurement errors in each $\ell$-band.}
\label{fig:diffexp}
\end{center}
\end{figure}

\subsection{Discrepancy between theoretical prediction and observation of the SZ power spectrum}
\label{sec:mismatch}

The discrepancy between the theoretical model predictions for the tSZ power spectrum and its experimental measurement were shown early on, following pure analytical modeling \citep{KS02,Bode09} and simulations \citep{Se10}. The tSZ power based on the \citetalias{Ar10} universal pressure model is not the exception, as is shown in Fig.~\ref{fig:diffexp}. The red solid curve is the prediction based on the \citetalias{Ar10} model, using Equations \ref{szpw_poisson} and \ref{A_pressure}, which has a factor of $\sim 2$ higher amplitude than the SPT measurement (marked by the gray-shaded region). Similar results were also shown by \citet{EM12}.

We point up that despite its high sensitivity, the SPT data is unable to constrain simultaneously both the amplitude and shape of the tSZ power spectrum. For this reason \citetalias{R12} used the \citet{Sh10} tSZ template to quote the amplitude at $\ell=3000$, which is $3.65\pm0.69~\mu$K$^2$. In Fig.~\ref{fig:diffexp} and subsequent plots, we show the SPT $68\%$ confidence region derived from the \citet{Sh10} model in gray bands, to provide a visual comparison with our best fit results. The plots are in terms of $D_{\ell}=\ell(\ell+1)C_{\ell}^{s}/2\pi$ with units of $[\mu$K]$^2$ at $150$~GHz. For quantifying the goodness of fit between a model prediction and the SPT data, we compute the probability to exceed (PTE, or the $p$-value) for the model using the actual CMB bandpower measurements from SPT, and not with the \citet{Sh10} model template. The PTE for the \citetalias{Ar10} model prediction is 0.0006, suggesting a very poor fit.

It is now possible to check the compatibility of the \plk and {\it Bolocam} measurements of the pressure profile with the SPT result following the procedure outlined in Section \ref{sec:method}. These are shown in Fig.~\ref{fig:diffexp} with the dotted and dashed lines, respectively. The prediction based on the mean \plk profile is very similar to the \citetalias{Ar10} model, whereas the mean {\it Bolocam} profile predicts higher amplitude for the tSZ power and returns further lower PTE value. This is primarily because of the shallower outer slope of the pressure profile as reported in the {\it Bolocam} paper, giving excess power al $\ell \lesssim 3000$. We take note of the fact that making a comparison with only the mean pressure profile, using the maximum likelihood values of the GNFW model parameters, is not correct since there is a large covariance between these parameters which will produce a range of equally likely pressure profiles. The \plk parameter covariance is not published, but we obtained the parameter chains for the GNFW model fit for {\it Bolocam} (J. Sayers, priv. comm.), which allow us to draw a 68\% credibility region of the tSZ power spectrum based on the {\it Bolocam} result. This is shown by the green shaded region in Fig.~\ref{fig:diffexp}, which is small compared to the roughly $5~\mu$K$^{2}$ difference between the predicted power and the SPT measured value. Based on similar parameter errors on the GNFW model fit by \citet{Pl13}, it can be assumed that the measured errors on the mean \plk profile cannot explain the mismatch with data either. 

It can be argued that the source of the discrepancy between the predicted amplitude of the tSZ power spectrum and its measurement comes from the assumed cosmological model. The tSZ power spectrum relies on the correct modeling of the halo mass function, but it has been proven that the halo mass function is known to an accuracy of about 5\% \citep{Ti08,Bh11}. However, the values of the key cosmological parameters differ significantly between the measurements made by different probes, like WMAP and \plk, which will result in large systematic changes in the tSZ power spectrum. In Section~\ref{sec:cosmology} we address this issue in more detail, and show if we ignore the large systematic difference between the WMAP and \plk cosmology, then the measurement uncertainties in any of the adopted set of parameters is not an issue. Moreover, use of the \plk cosmology increases the predicted tSZ power amplitude by roughly factor 2, thereby making the tension between theory and observation more severe. Since we observe a reduced amplitude of the tSZ power, the most likely explanation for that lower amplitude must be astrophysical, and our use of WMAP cosmology amounts to a more conservative modification of the ICM pressure distribution to match theory and observation.

\begin{figure}
\begin{center}
\includegraphics[width=\columnwidth]{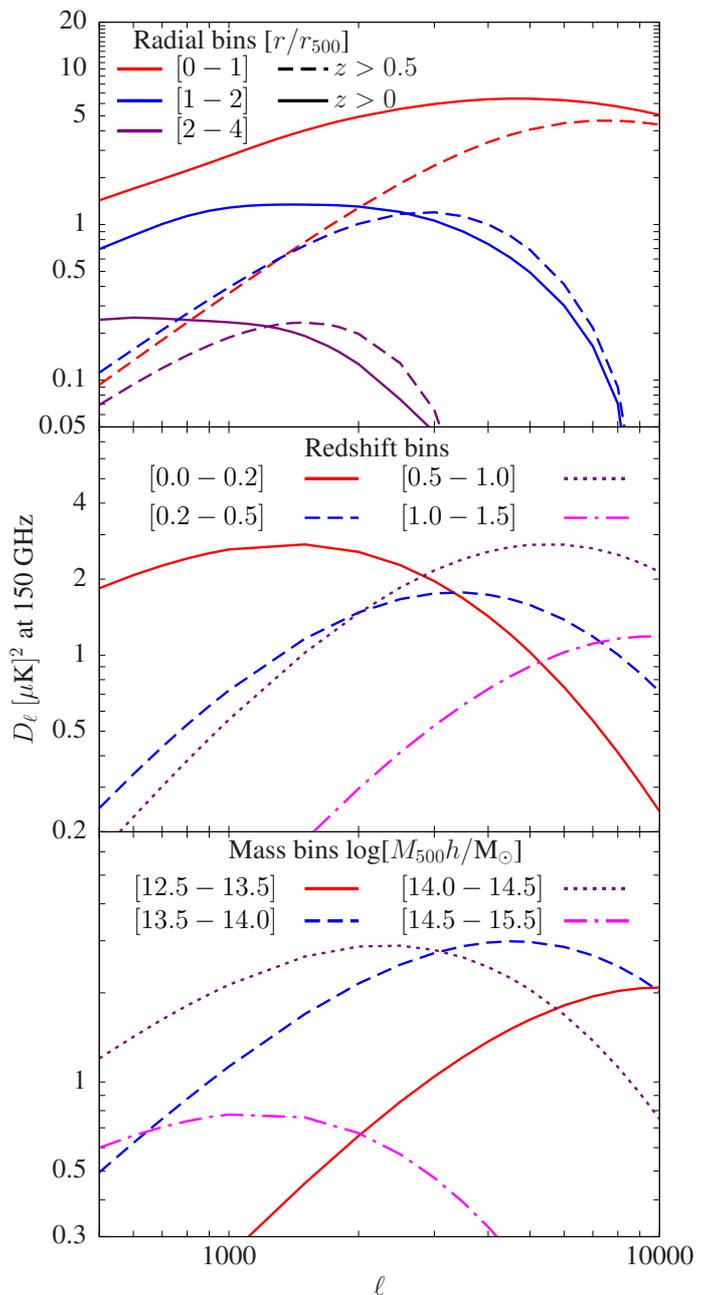}
\caption{Contribution of the tSZ power spectrum in different cluster radial, mass and redshift bins. The plot is in terms of $D_{\ell}=\ell(\ell+1)C^{s}_{\ell} / 2\pi$ with units of $[\mu$K]$^2$ at $150$~GHz. For these illustrative plots we have used the \citetalias{Ar10} pressure profile without modifications, and factored in the contribution from intrinsic scatter. The numbers in the square brackets mark the radial, redshift and mass bins for the individual curves. In the upper panel, the dashed lines only show the contribution from high-redshift ($z>0.5$) clusters. The anisotropy power from cluster outskirts ($r>r_{500}$) becomes increasingly important at $\ell \lesssim 3000$.}
\label{fig:diffmz-ps}
\end{center}
\end{figure}

\subsection{Radial, mass and redshift contribution to the power spectrum}

We split the tSZ power into mass, redshift and radial bins to identify where the main contributions to the tSZ power come from. The results in this Section are similar to those already presented in \citet{KS02}, \citet{Ba12b} and \citet{McC14}, although we use the universal pressure model of \citetalias{Ar10} for our analysis. 

First, we radially truncate the pressure profile to investigate which regions of the galaxy clusters contribute more to the total tSZ power spectrum. The differential contributions from three radial bins are shown in Fig.~\ref{fig:diffmz-ps} (top panel). Earlier works have shown the results for radial contributions in cumulative plots, which automatically include the cross-correlation of the tSZ power between different bins. We make a differential plot for ease of comparison and take the cross-correlation terms into account. When considering all clusters at all redshifts, most of the power ($\sim$ 85\%) on small angular scales ($\ell>3000$) comes from $r<r_{500}$, since bulk of the cluster tSZ signal comes from this central region. Outskirts of galaxy clusters play an important role at lower $\ell$. The tSZ power increases by $\sim 50\%$ at $\ell=500$ when the upper limit of the radial integration is increased to $2r_{500}$. Thereafter, the tSZ amplitude does not vary much if the integration limit is extended to radii $r>4r_{500}$. If we only consider clusters at high redshifts ($z > 0.5$), the power contribution from the outskirts will be roughly equal to that coming from the inner region for $\ell \lesssim 2000$ (dashed-lines in the upper panel of Fig.~\ref{fig:diffmz-ps}).

In a similar way, we calculate the tSZ power spectrum in mass and redshift bins, which are shown in the middle and bottom panels of Fig.~\ref{fig:diffmz-ps}, respectively. On large angular scales ($\ell<2000$) the largest contribution to the tSZ power comes from high mass ($14<\log[M_{500}h/$~M$_{\odot }]<14.5$) and low redshift objects ($z<0.2$). Above $\ell \sim 3000$, the tSZ power is dominated by low mass ($13.5<\log[M_{500}h/$~M$_{\odot }]<14$) and high redshift ($z>0.5$) galaxy groups/clusters. The very massive objects ($\log[M_{500}h/$~M$_{\odot }]>14.5$) have a negligible contribution on small angular scales. Therefore the clusters that are currently constrained from direct tSZ observation by \plk and {\it Bolocam} are not representative in terms of the measurement of the tSZ power spectrum. Objects with very low mass, or redshift $z>1$ dominate the tSZ power only at multipoles larger than $\ell \sim 15000$, assuming the extrapolation of the \citetalias{Ar10} pressure model is correct in such extreme cases. 

From the illustrations above it can be seen that the tSZ power near the angular scales of its expected peak is dominated by contributions from the low mass clusters or groups at intermediate redshifts, for which there are little observational constraints on their ICM properties. Thus, using the fiducial values for the \citetalias{Ar10} pressure model on high redshift galaxy groups and clusters could be the source of the over-estimation of the tSZ power. Likewise, the \citetalias{Ar10} pressure profile beyond $r_{500}$ was constrained from hydrodynamical simulations, which could also be overestimating the thermal pressure component in the outskirts, giving more tSZ power. Thus, two obvious choices for modifying the \citetalias{Ar10} pressure profile would be {\it i)} decreasing the pressure amplitude with redshift that offsets the self-similar evolution, or {\it ii)} decreasing the thermal pressure support in the outskirts. 

The mass dependence of the pressure normalization (or the outer pressure slope), as discussed in Section \ref{sec:results}, are generally better constrained from observation (or simulations) and are not the main focus of the current paper.

\subsection{Effect of cluster morphology}

Before proceeding to modify the universal pressure profile, it is natural to ask whether an over-abundance of merging systems at high redshifts can be responsible for the lower measured tSZ power. This follows from the result of \citetalias{Ar10} who found, with high significance based on X-ray data, that disturbed clusters have lower pressure near the core region compared to the mean. In the standard $\Lambda$CDM scenario the number of mergers within a time interval is a slowly increasing function of halo redshift and mass \citep[e.g.,][]{Fa10}, and there is some evidence of this increased merger fraction from X-ray selected clusters \citep{Mau12,ME12}. The resulting change in the cool-core (CC) to non cool-core (NCC) cluster ratio with redshift might be causing the lower amplitude of the tSZ power spectrum in the SPT data.

\begin{figure}
\begin{center}
\includegraphics[width=\columnwidth]{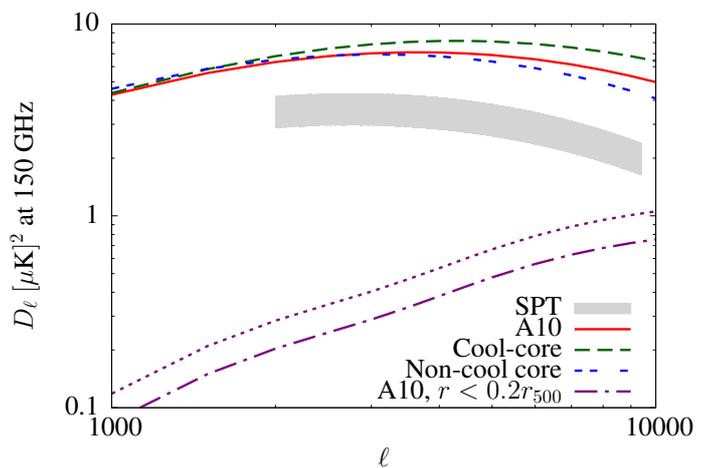}
\caption{Predictions for the tSZ power spectrum for cluster morphological evolution. The red-solid line is given by the mean \citetalias{Ar10} pressure profile, the green-dashed line for cool-core clusters, and the blue short-dashed line for non-cool clusters. The purple dot-dashed line shows the relative contribution from the core regions of galaxy clusters ($r < 0.2 r_{500}$) to the total tSZ power, and the dotted line above it factors in an additional scatter contribution (40\%) for the core region. This plot shows that, unlike the X-ray luminosity, the core region contributes very little to the tSZ anisotropy power.} 
\label{fig:coreregion}
\end{center}
\end{figure}

\citetalias{Ar10} divided their cluster sample into CC and NCC clusters and provided parametric fits for the respective populations \citepalias[see Appendix~C of][]{Ar10}. The CC clusters have more peaked profile at the center, the region that produces bulk of the emissivity in X-rays. However, from Fig.~\ref{fig:coreregion} we can conclude that the core region ($r<0.2r_{500}$) contributes a negligible fraction of the tSZ power on scales larger than $\sim 1$~arcmin: the contribution is only $5\%$ at $\ell=3000$, and rises up to nearly $17\%$ at $\ell=10000$. 
\citetalias{Ar10} found the pressure profiles of the CC and NCC cluster samples nearly self-similar in the intermediate ($r\sim r_{500}$) to outer regions, and currently there are no direct evidence for a systematic difference between the CC and NCC cluster pressure distribution from tSZ data. Given that restriction, the CC and NCC clusters produce roughly the same result for tSZ power (Fig.~\ref{fig:coreregion}). We thus conclude that an increased occurrence of NCC clusters at high-$z$ cannot be the explanation of the low measured value of the tSZ power, if those NCC clusters follow the same mass and redshift scaling as given for the universal pressure profile by \citetalias{Ar10}. 

%%%%%%%%%%%%%%%%%%%%%%%%%%%%%%%%%%%%%%%%%%%%%%%%%%%%%%%%%%%%%%%%%%%%%%%%%%%%%%%%%%%%%%%%%%%%%%%%%%%%%%%%%%%%%%%%%%%%%%%%%%%%%%%%%%%%%%%%%%%%%%%

\section{Results}
\label{sec:results}

This Section presents our main results, following various attempts at modifying the universal pressure model. We group these model changes according to their deviations from a simple self-similar scaling.

\subsection{Modification following strictly self-similar evolution}

In the classical self-similar scenario of cluster evolution, the baryon distribution will have the same shape and amplitude once they are scaled to the cluster mass and redshift with the standard scaling powers \citep[e.g.,][]{Boe12}. In context of the \citetalias{Ar10} pressure profile, this means that clusters at all mass and redshift will have the same set of amplitude and shape parameters: $\{P_0, c_{500}, \alpha, \beta, \gamma\}$, and the total pressure amplitude will scale with redshift as $E(z)^{8/3}$. In a first attempt to keep the redshift evolution unchanged, we try to find a suitable set of shape parameters that will remove the tension between the model predictions and SPT data.

\begin{figure}
\begin{center}
\includegraphics[width=\columnwidth]{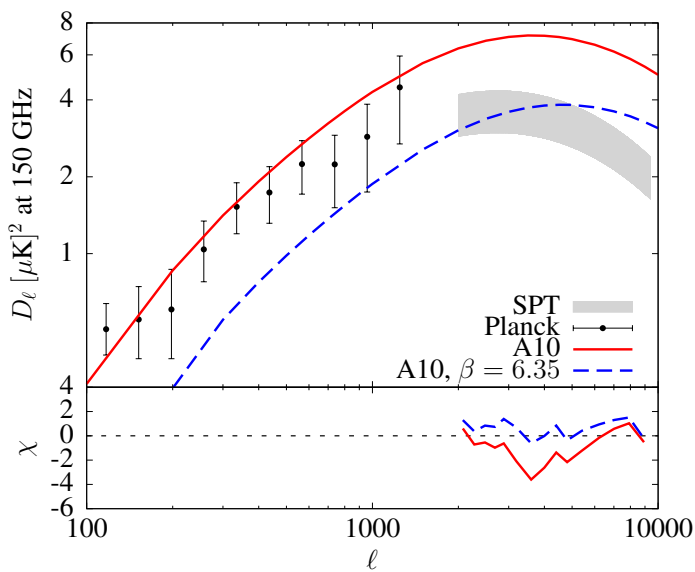}
\caption{Predictions for the tSZ power spectrum for self-similar evolution. The red solid line is the tSZ power spectrum given by the \citetalias{Ar10} pressure profile (Eq.~\ref{A_pressure}). The shaded gray region represent the 1$\sigma$ constraints from the SPT based on the \citet{Sh10} template (see text for details). The blue dashed line represents the GNFW model with our best fit outer slope ($\beta=6.35$), which provides good fit to the actual SPT data, as shown by the $\chi$ plot in the bottom panel. The black data points are the marginalized bandpowers of the Planck tSZ power spectrum, taken from \citet{Pl13b}.}
\label{fig:diffbeta_Planck}
\end{center}
\end{figure}

\subsubsection{Constraints on the outer slope of the pressure profile}
\label{sec:outslope}

As mentioned previously, the ``universal'' pressure profile was constrained from X-ray observations out to radii $r\sim r_{500}$, and extrapolated beyond $r_{500}$ to match hydrodynamical simulation results. The outer slope parameter is denoted by $\beta$, whose value is fixed at $\beta=5.49$ in the \citetalias{Ar10} paper. A significant amount of the tSZ signal comes from $r>r_{500}$, more than 50\% if we neglect the few nearby, high-mass clusters (that are generally resolved in deep tSZ surveys), therefore the impact of $\beta$ on the tSZ power amplitude is pivotal. A higher value of $\beta$ would imply less thermal pressure. Physical reason for lower thermal pressure can be additional pressure support from gas bulk motions, usually triggered by infalling or merging sub-halos. Recent results from numerical simulations \citep{Na07,Lau09,Ne14} as well as analytical modeling \citep{Shi14} have shown that this non-thermal pressure contribution is small in the inner regions, but rapidly increases with radius. We therefore concentrate on modifying $\beta$, given the least amount of observational constraint on its value, while keeping the redshift evolution of the pressure amplitude at $E(z)^{8/3}$ as in Eq.~\ref{A_pressure}.

The best fit value obtained from our MCMC sampling is $\beta=6.35\pm0.19$, together with the CIB contribution terms $D_{3000}^{\textrm{P}}=7.58\pm0.28$ and $D_{3000}^{\textrm{C}}=6.42\pm0.54$. The resulting PTE$=0.80$ suggests a good fit to the CMB bandpower data. This value of $\beta$ is considerably higher than the one assumed by \citetalias{Ar10} ($\beta=5.49$), implying a much lower thermal pressure support in the outskirts. This new value of $\beta$ has very little effect on the inner pressure profile ($< 1\%$ at $r<<r_{500}$), but reduces the pressure amplitude by $\sim 40\%$ at $r_{500}$. The effect is significant on the power spectrum after projection, especially on large scales, where the new tSZ power amplitude is lowered by $\sim 50\%$ compared to the \citetalias{Ar10} values (Fig.~\ref{fig:diffbeta_Planck}, blue dashed line). Furthermore, the peak of the tSZ power spectrum is shifted to smaller angular scales, near $\ell\sim 4500$. We note that the shape of the tSZ power spectrum differs strongly compared to the \citet{Sh10} template (gray band), but the new shape provides acceptable fit to the SPT data together with the above CIB power amplitudes.

\subsubsection{Possible tension with \plk and {\it Bolocam} results}

\begin{figure}
\begin{center}
\includegraphics[width=\columnwidth]{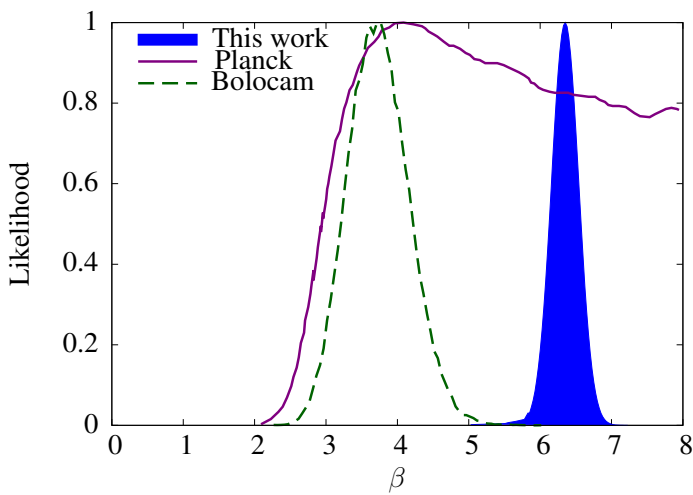}
\caption{1D likelihood curves for outer slope parameter $\beta$. The purple solid line shows the constraints from \plk. Results from {\it Bolocam} are shown by the green dashed line. The blue-filled curve represents our best fit constraint on $\beta$.}
\label{diff:likelihoods}
\end{center}
\end{figure}

\noindent The marginalized value of the outer pressure profile slope, $\beta=6.35\pm0.19$ with 68\% confidence, is higher than the mean values obtained from direct cluster SZ profile measurement by {\it Bolocam} and \plk experiments. However, in a GNFW model fit the value of $\beta$ generally highly degenerates with other parameters, in particular it anti-correlates with the scale radius (or equivalently, $c_{500}$), as shown by \citet{Pl13}. Therefore we must compare our best fit values with the marginalized errors on $\beta$ from other experiments. This is shown in Fig.~\ref{diff:likelihoods}, where we plot the normalized likelihood distributions for the outer slope parameter $\beta$ from \plk and {\it Bolocam} fits. As can be seen, there is significant tension for such a steep value of outer slope with {\it Bolocam} data, whereas it is consistent with the \plk measurement. A possible cause can be that the {\it Bolocam} team fixes the gas concentration parameter $c_{500}$, which restricts their likelihood range of $\beta$, even though both our modeling and the {\it Bolocam} work by \citet{Sa13} use the same fixed value $c_{500} = 1.18$ from the \citetalias{Ar10} model. The sensitivity of the \plk measurement to the slope parameters is possibly lower due to its large beam, except for a few nearby clusters.

A general consequence of having a steeper outer slope is that it will inevitably reduce the tSZ power at low $\ell$ values ($\ell < 1000$), as seen in subsection~\ref{sec:outslope}. This then leads to some tension with the tSZ power measurements based on \plk data \citep{Pl13b}, which we show in Fig.~\ref{fig:diffbeta_Planck}. The \plk marginalized bandpower values are taken directly from the \plk collaboration paper. Without a knowledge of the covariance we cannot compute the $\chi^2$ or the PTE value of our model from \plk data, but a clear tension can be seen from the figure.

\subsection{Weakly self-similar: changing pressure normalization with redshift or mass}

It is possible to imagine scenarios where the amplitude of the pressure distribution in galaxy clusters deviate from a strictly self-similar evolution, i.e., not scaling as $P(r) \propto M_{500}^{2/3}$ or/and $P(r) \propto E(z)^{8/3}$. In fact \citetalias{Ar10} already show that the mass scaling of the pressure profile is not strictly self-similar, there is an additional factor, $\alpha_P=0.12$, in the mass-scaling power (see Eq.~\ref{A_pressure}). The reason behind this deviation from self-similarity is the empirical calibration of the \citetalias{Ar10} pressure profile against the measured $Y_X - M_{500}$ scaling of \citet{Ar07}, which was based on a subset of REXCESS clusters at low redshifts. Therefore, while the small deviation in the mass exponent in Eq.~\ref{A_pressure} is well-measured in the local universe, its redshift dependence remains largely unexplored. 

In this Section we explore scenarios where the redshift evolution of the pressure amplitude deviates from the $E(z)^{8/3}$ scaling, while keeping the shape (Eq.~\ref{gnfw_profile}) constant. Physical motivation for such redshift dependence can be found from the observed scaling of the $L_X-T_X$ scaling relation \citep[e.g.,][]{Rei11}, possibly relating to a gas mass fraction, $f_{\gas}$, evolution in groups and clusters which we discuss subsequently. As an extension to this model, we consider cases where the redshift evolution depends also on mass, in line with the observed difference in the mass dependence of $f_{\gas}$ in groups and clusters.

\subsubsection{Departure from self-similar redshift evolution}
\label{subsec:nonSSz}

As discussed in Section~\ref{subsec:ssmodel}, in the self-similar model the redshift evolution of the global properties of galaxy clusters is described by a simple power law of the $E(z)$ parameter. However, non-gravitational processes, like cooling and feedback, can alter the expected redshift evolution parametrization \citep[e.g.,][]{Vo05}. Such possible deviations from self-similarity can be considered through a $(1+z)$ term lacking a better understanding of their origin. Recent semi-analytical works \citep{Sh10,Ba12a} have also used a power-law dependence of $(1+z)$ for modeling the non-thermal pressure evolution.

Following the above argument, and keeping the functional form of the \citetalias{Ar10} pressure profile unchanged at low redshifts, we introduce an additional $(1+z)$ dependence of the form 
\begin{equation}
 P_{e}(r)\propto E^{\frac{8}{3}}(z)(1+z)^{\alpha_{z}}M_{500}^{\frac{2}{3}+\alpha_{p}}.
\label{pe_alphaz}
\end{equation}

\begin{figure}
\begin{center}
\includegraphics[width=\columnwidth]{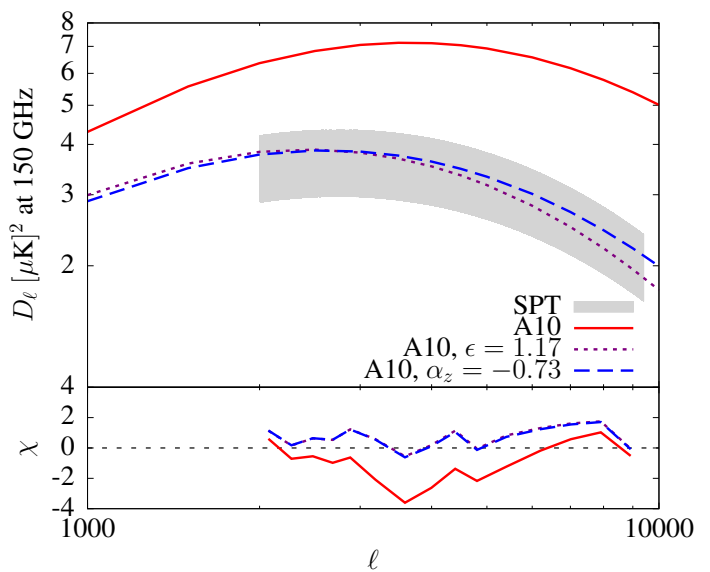}
\caption{Predictions for the tSZ power spectrum for a weak departure from self-similarity, affecting only the pressure normalization. The red solid line is the tSZ power spectrum given by the \citetalias{Ar10} pressure model, and the shaded gray region is the 1$\sigma$ scatter around the best fit Shaw model with SPT data. The blue dashed line represents the \citetalias{Ar10} model with $\epsilon=1.17$ (Eq.~\ref{A10_mod}). The purple dotted line represents \citetalias{Ar10} model with $\alpha_{z}=-0.73$ (Eq.~\ref{pe_alphaz}). Both parameters, $\epsilon$ and $\alpha_{z}$, modify the redshift evolution of the pressure profile to reduce its amplitude at high-$z$.}
\label{fig:tszpw_evolution}
\end{center}
\end{figure}

The parameter $\alpha_{z}$ signifies the departure from the self-similar evolution, and we constrain its value by comparing with the SPT measurements \citepalias{R12} through our MCMC method. The best fit value is $\alpha_{z}=-0.73\pm0.16$, $D_{3000}^{\textrm{P}}=7.69\pm0.27$, and $D_{3000}^{\textrm{C}}=6.35\pm0.49$, with a PTE of 0.78. The overall effect of such non self-similar evolution is to lower the amplitude of the tSZ power spectrum (purple-dotted line in Fig.~\ref{fig:tszpw_evolution}), since as the negative value for $\alpha_{z}$ implies, the pressure amplitude in groups/clusters decreases with increasing redshift. The modified shape of the tSZ power spectrum is in good agreement with the Shaw et al. template (Fig. \ref{fig:tszpw_evolution}).

From a cosmological point of view, a power-law dependence of $E(z)$ is a more attractive parametrization for the non-self-similar evolution. \citet{EM12} have proposed such a model by introducing the parameter $\epsilon$ into the $E(z)$ power for the pressure scaling:
\begin{equation}
P_{e}(r)\propto E(z)^{\frac{8}{3}-\epsilon}M_{500}^{\frac{2}{3}+\alpha_{p}}.
\label{A10_mod}
\end{equation}
As before, we can constrain the value of $\epsilon$ by comparing with the SPT measurements. We find values of $\epsilon=1.17\pm0.27$, $D_{3000}^{\textrm{P}}=7.69\pm0.26$, and $D_{3000}^{\textrm{C}}=6.40\pm0.51$, with a resulting PTE of 0.79 (blue-dashed line in Fig.~\ref{fig:tszpw_evolution}). We cannot directly compare with the results of \citet{EM12}, since they constrained their $\epsilon$ value by comparing with simulated tSZ power spectrum templates \citep{Ba10,Tr11}, and incorporate another additional normalization parameter ($A$) for their model that should be highly degenerate with the redshift evolution term $\epsilon$.

Depending on the physical origin of the non-self-similar evolution, either an $E(z)$ or a $(1+z)$ power-law dependence will describe the pressure profile modification correctly. Since both these parametrizations result in similar changes to the $P(r)$, and show similar relative errors, we have opted for keeping both cases in our analyses.

\subsubsection{Association with $f_\gas$ and X-ray scaling laws}
\label{subsec:fgasMdep}

The pressure-mass, $P-M$, scaling relation has a direct dependence on the gas mass fraction, $f_{\textrm{gas}}$: $P\propto f_{\textrm{gas}}E^{8/3}M^{2/3}$. This quantity is usually assumed as constant, and the \citetalias{Ar10} work make no explicit statement on gas mass fraction either. However, the $P-M$ relation used by \citetalias{Ar10} deviates from the self-similar prediction, having a slightly stronger mass dependence: $P\propto E^{8/3}M^{2/3+0.12}$. By comparing the \citetalias{Ar10} $P-M$ relation with the self-similar prediction, we can assume that this excess is the result of the gas mass fraction: $f_{\textrm{gas}}\propto M^{0.12}$. This is motivated by studies which have found that $f_{\textrm{gas}}$ increases with mass of galaxy groups and clusters \citep{Vi06,Ar07,Pra09,Su09}, since non-gravitational processes (e.g., AGN feedback and star formation) produce a larger impact on galaxy groups than in clusters. \citet{Pra09}, for example, found a relatively strong mass dependence of $f_{\textrm{gas}}$ in the REXCESS sample: $f_{\textrm{gas}}\propto M^{0.2}$; and for low mass regime, \citet{Su09} constrained the mass dependence in the range: $f_{\textrm{gas}}\propto M^{0.16-0.22}$.

\begin{figure}
\begin{center}
\includegraphics[width=\columnwidth]{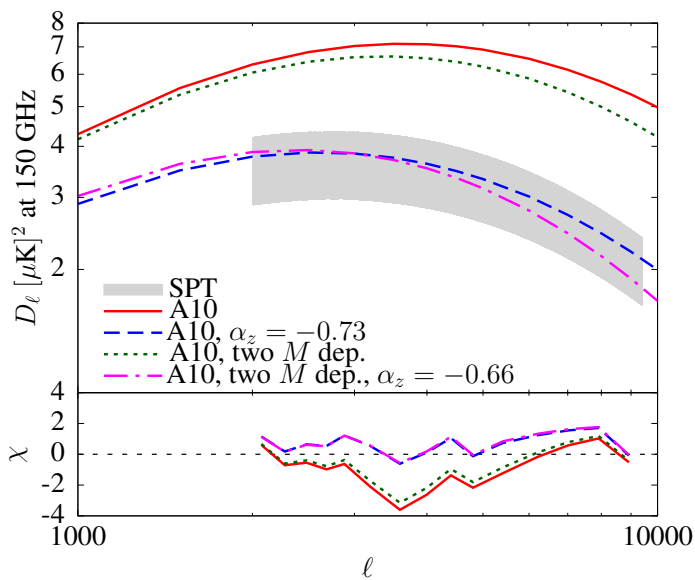}
\caption{Predictions for the tSZ power spectrum for a weak departure from self-similarity, factoring in additional mass-dependence for the scaling. The red solid line and the gray shaded regions have the same meaning as in earlier figures. The blue dashed line represents the tSZ power spectrum with $\alpha_{z}=-0.73$ (Eq.~\ref{pe_alphaz}), as in Fig. \ref{fig:tszpw_evolution}. The green dotted line is given by a modified \citetalias{Ar10} pressure profile that has double mass dependence: $f_{\textrm{gas}}\propto M^{0.2}$ for masses below $M_{500}=10^{14}~h^{-1}$~M$_{\odot}$, and $f_{\textrm{gas}}\propto M^{0.12}$ above this mass limit. Since this change alone is not enough to reach the SPT constraints, we need to introduce a redshift evolution, the result of which is shown by the magenta dash-dotted line.}
\label{fig:fgasmod}
\end{center}
\end{figure}

In order to assess the impact of different mass dependence of $f_{\textrm{gas}}$ on the tSZ power spectrum, we introduce two distinct power laws for the mass dependence in the \citetalias{Ar10} pressure profile (Eq.~\ref{A_pressure}): $f_{\textrm{gas}}\propto M^{0.2}$ and $f_{\textrm{gas}}\propto M^{0.12}$ for masses below and above $M_{500}=10^{14}~h^{-1}$~M$_{\odot}$, respectively. Fig.~\ref{fig:fgasmod} shows the small effect of this broken power law for mass dependence on the tSZ power spectrum (green-dotted line), which is not enough to explain the discrepancy between tSZ power spectrum predictions and the SPT constraints. This shows that a small departure from the self-similar mass scaling, in accordance with the observational results, is not sufficient to explain the low amplitude of the tSZ power spectrum by itself; one needs to consider a modification to the redshift evolution as well.
By using our MCMC method, we found that the necessary evolution is given by: $\alpha_{z}=-0.66\pm0.15$, $D_{3000}^{\textrm{P}}=7.70\pm0.21$, and $D_{3000}^{\textrm{C}}=6.38\pm0.47$, with a PTE of 0.79 (shown in Fig.~\ref{fig:tszpw_evolution} by the magenta dash-dotted line).

\subsubsection{$f_{\gas}$ evolution vs. X-ray data and simulations}

In the previous subsection we assumed that the weakly non-self-similar $P-M$ scaling used by \citetalias{Ar10} is due to the fact that $f_{\textrm{gas}}$ has a mass dependence. In a similar manner, the non-self-similar evolution required to explain the discrepancy between the SPT measurements and the theoretical predictions of the tSZ power spectrum (subsection~\ref{subsec:nonSSz}) can be attributed to an evolution in $f_{\textrm{gas}}$. This assumption is motivated by recent observations that show scaling relations, which have a direct dependency on the $f_{\textrm{gas}}$, do not always follow a self-similar evolution \citep[see discussion in][]{Boe12}. For example, \citet{Rei11} and \citet{Hi12} have measured the $L_X-T_X$ scaling relation in different redshift ranges, and they find it to be non-self-similar.

The X-ray bolometric luminosity scales with $f_{\gas}$ as
\begin{equation}
 L_{\textrm{bol}}\propto f_{\textrm{gas}}^{2}T^{2}E(z).
\end{equation}
If we assume that the temperature scaling remains self-similar, this would suggest an evolving baryon fraction in clusters. Thus, our tSZ power spectrum based on an evolving $f_{\textrm{gas}}$ model, following the results from previous subsection \ref{subsec:fgasMdep}, would suggest
\begin{equation}
 E^{-1}L_{\textrm{bol}}\propto T^{2.36}(1+z)^{-1.82\pm0.302},
\end{equation}
for the $(1+z)^{\alpha_z}$ scaling, or 
\begin{equation}
L_{\textrm{bol}}\propto T^{2.36}E^{-1.58\pm0.54},
\end{equation}
for the $E(z)^{8/3-\epsilon}$ scaling. We compare these results with the XCS cluster sample result from \citet{Hi12}, and also from \citet{Rei11} who use an ad-hoc high-$z$ cluster sample. \citet{Hi12} have found the scaling for the bolometric luminosity as $E^{-1}L_{\textrm{bol}}\propto T^{3.18}(1+z)^{-1.7\pm0.4}$
or $L_{\textrm{bol}}\propto T^{3.18}E^{-1.2\pm0.5}$. The result from \citet{Rei11} is a less significant change with redshift for the soft-band luminosity: $L_{X}\propto T^{2.70\pm0.24}E^{-0.23^{+0.12}_{-0.62}}$. We see that our results are generally consistent with those from \citet{Hi12}, but there is disagreement with the \citet{Rei11} scaling. What is significant, however, is that the errors on the redshift evolution term from our modeling are similar to those available at present from direct X-ray observations. This illustrates the promise of tSZ power spectrum measurements to constrain cluster scaling relations, and we shall discuss its future prospects in subsection \ref{sec:ccat}.

\begin{figure}
\begin{center}
\includegraphics[width=\columnwidth]{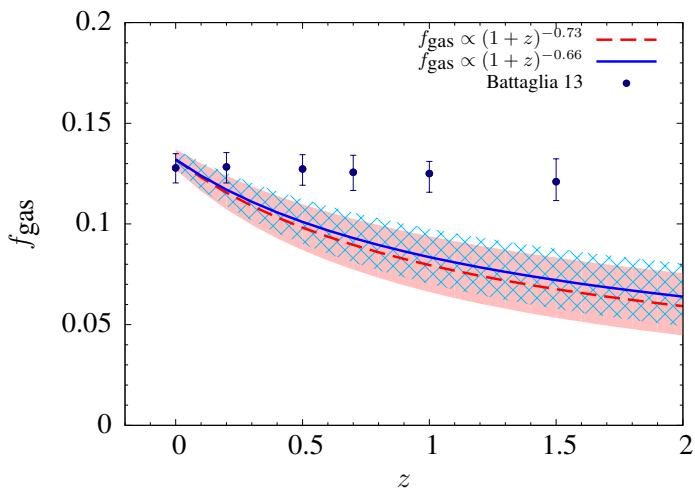}
\caption{Redshift evolution of the gas mass fraction, $f_{\textrm{gas}}$, based on our modeling. The blue solid line is for redshift-dependent $f_{\gas}$ following the $(1+z)^{\alpha_z}$ power law, and the red dashed line if the same but with two different mass scaling (see subsection \ref{subsec:fgasMdep}). The hatched and shaded regions mark the $1\sigma$ confidence intervals around these lines, respectively. Points with error bars are taken from \citet{Ba13} simulations, who compute the mean $f_{\textrm{gas}}$ within $r_{200}$.}
\label{fig:fgas_diffevol}
\end{center}
\end{figure}

Currently there are no direct $f_{\gas}$ measurement in a mass-limited cluster sample out to high redshifts. The works by \citet{Allen04,Allen08} use carefully selected relaxed clusters from X-ray survey data, where they find that $f_{\gas}$ remains practically unchanged with redshift. The results from complete X-ray samples are restricted to low redshifts, for example the REXCESS sample by \citet{Pra09}. Therefore, we make a comparison with recent results from N-body hydrodynamical simulations of clusters that have aimed at measuring the evolution of the baryonic component. Fig.~\ref{fig:fgas_diffevol} shows a comparison of the $f_{\gas}$ from \citet{Ba13}, who use hydrodynamical TreePM-SPH simulations including cooling, star-formation, and supernova and AGN feedback, with our results. Clearly both our single power law evolution and the broken mass-dependent evolution models are inconsistent with \citet{Ba13} predictions, which show a negligible change in $f_{\gas}$ with redshift.

\subsection{Non-self-similar: an evolving shape of the pressure profile}
\label{sec:outer_slope}

In our study of a deviation from self-similarity, we have assumed the shape of the pressure profile remains constant with redshift, such that the outer pressure slope parameter $\beta$ does not change with $z$. In reality this is unlikely to be true, since the cluster merger fraction steadily increases with redshift, meaning departure from hydrostatic equilibrium should become more significant at high-$z$, making non-thermal pressure support more prominent. \citet{Sh10} considered this effect and identified a redshift evolution of the non-thermal pressure support as potentially the most significant contributor to the lower amplitude of the tSZ power spectrum. An enhancement of the non-thermal pressure (random gas motions) with redshift is also shown by recent hydrodynamical simulations \citep{Lau09,Ne14}, who in addition find that there is practically no mass dependence for this effect. Our treatment of a non-self-similar pressure shape, therefore, only consists of an evolution with redshift and no scaling with cluster mass.

\begin{figure}
\begin{center}
\includegraphics[viewport= 65 28 386 332,clip,width=\columnwidth]{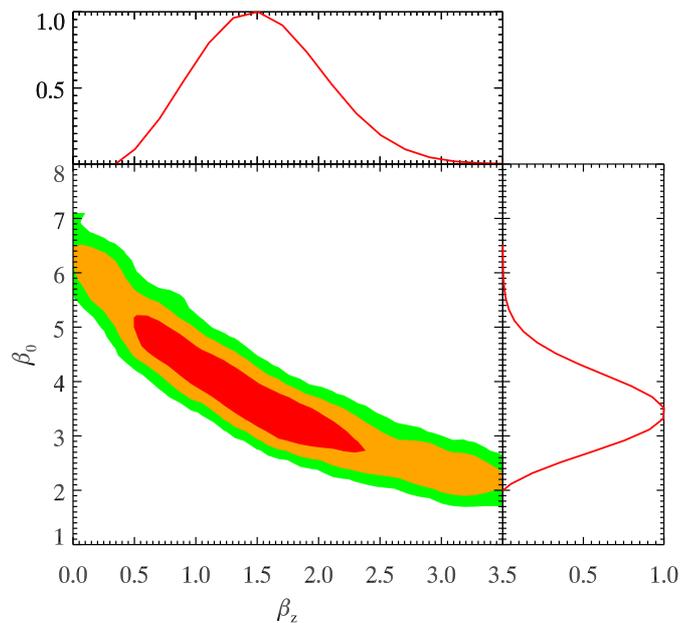}
\caption{2D Likelihood contours for the correlation between the $\beta_{0}$ (the outer slope parameter at $z=0$) and $\beta_{z}$ (its redshift evolution, see Eq.~\ref{eq:betaevol}). The colored contours show the 1, 2, and 3$\sigma$ constraints, and the marginalized values are shown in the side panels. Both $\beta_{0}$ and $\beta_{z}$ tend to lower the tSZ power amplitude and hence anti-correlate.}
\label{fig:2beta_degeneracy}
\end{center}
\end{figure}

We consider a model of this redshift-dependent steepening of the pressure profile using a simple, analytic form for the slope parameter $\beta$ as follows:
\begin{equation}
\beta = \beta_0 (1+z)^{\beta_z},
\label{eq:betaevol}
\end{equation}
Here $\beta_0$ is the outer slope parameter at $z=0$, roughly reminiscent of the low-redshift measurements by \citetalias{Ar10} and Sun et al. (2011), and $\beta_z$ is its redshift scaling. Fig.~\ref{fig:2beta_degeneracy} shows the result of model constraints using this new redshift-dependent term. The parameter $\beta_z$ is highly degenerate with $\beta_0$, with large errors on their marginalized values: $\beta_z=1.50^{+0.60}_{-0.55}$ and $\beta_0=3.50^{+0.80}_{-0.70}$ with a PTE$=0.79$. Likewise, if we parametrize the redshift evolution of $\beta$ with an $E(z)$ power-law as
\begin{equation}
\beta = \beta_0^{\prime} E(z)^{\beta_z^{\prime}},
\end{equation}
we obtain $\beta_z^{\prime}=2.12^{+1.19}_{-1.07}$ and $\beta_0^{\prime}=3.79^{+0.79}_{-0.71}$ with a PTE$=0.76$. This parameter degeneracy is a general conclusion whenever we try to constrain a pressure profile model parameter and its redshift evolution simultaneously from the current SPT data.

%%%%%%%%%%%%%%%%%%%%%%%%%%%%%%%%%%%%%%%%%%%%%%%%%%%%%%%%%%%%%%%%%%%%%%%%%%%%%%%%%%%%%%%%%%%%%%%%%%%%%%%%%%%%%%%%%%%%%%%%%%%%%%%%%%%%%%%%%

\section{Discussion and outlook}
\label{sec:discussion}

In this Section, we discuss the limitations of the present generation tSZ power spectrum experiments to constrain multiple model parameters for the ICM pressure. We then make predictions for future experiments using simulated bandpower data, based on the same SPT baseline model but scaled to the expected sensitivities for those new tSZ surveys. Finally, we consider the impact of cosmological parameter uncertainties on our methodology of constraining the ICM pressure from current and future tSZ power measurements.

\subsection{Need for better tSZ power spectrum measurements}

In the course of our modification attempts for the ICM pressure profile from its universal shape and amplitude, we found several potential solutions that can bring the power spectrum amplitude in accordance with the SPT data, but none of these solutions are fully satisfactory in light of the current data or simulations. Evidently, more than one effect is responsible for the observed low tSZ power, such as a combination of steeper pressure profile in the cluster outskirts (and possibly its redshift evolution) and a redshift-dependent baryonic fraction. Unfortunately, current ground- or space-based tSZ experiments do not have the requisite sensitivity and resolution to simultaneously constrain both the shape and the amplitude of the tSZ power spectrum, while separating it from the other multiple astrophysical components affecting the CMB bandpowers at 150 GHz. 

\begin{figure}
\begin{center}
\includegraphics[viewport= 60 22 443 332,clip, width=\columnwidth]{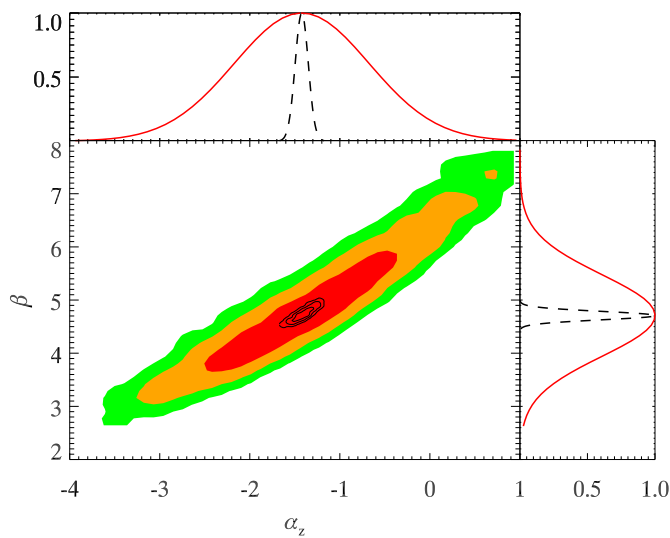}
\caption{2D likelihood contours for the $\beta$ and $\alpha_{z}$ parameters and their marginalized values. The colored contours show the 1, 2, and 3$\sigma$ constraints available from the SPT bandpowers of \citetalias{R12}. The black solid lines show the expected constraints from the CCAT tSZ survey. The marginalized errors for CCAT (dashed lines) are almost an order of magnitude smaller.}
\label{fig:predictions_SPTCCAT}
\end{center}
\end{figure}

As an illustration we pick up the two most prominent parameters featured in our analysis: the slope parameter $\beta$ and the non-self-similar term $\alpha_{z}$, to demonstrate this parameter degeneracy. Results are shown in Fig.~\ref{fig:predictions_SPTCCAT} with colored contours, and marginalized values in red solid lines. When both parameters are varied simultaneously we obtain $\alpha_{z}=-1.42\pm0.75$ and $\beta=4.71\pm0.71$. Clearly, none of these constraints are very informative, the non-self-similar evolution term is consistent with zero at $2\sigma$. A similar case arises for any other parameter combination that can each individually attribute to the lower tSZ power measurement. 

Upcoming SZ survey experiments, however, will have sufficient sensitivity and sky-coverage to place simultaneous constraints on the amplitude and the shape of the tSZ power spectrum. This will bring in a significant improvement in the parameter uncertainties (e.g., $\beta$ or $\alpha_z$), and help to break the current parameter degeneracies. Two such experiments are CCAT\footnote{www.ccatobservatory.org} and SPT-3G. In the following we use CCAT to demonstrate the improved parameter constraints from future SZ experiments.

\subsection{Predictions for CCAT}
\label{sec:ccat}

\begin{figure}
\begin{center}
\includegraphics[width=\columnwidth]{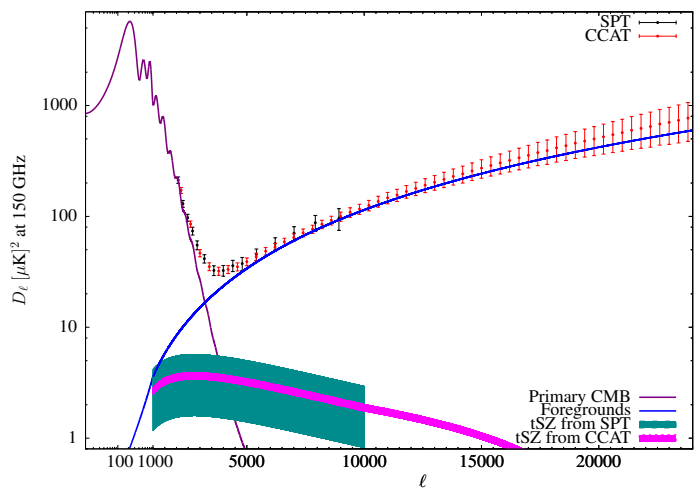}
\caption{The current SPT bandpower measurements for the total CMB anisotropies (black data points, from \citetalias{R12}), and the predicted bandpowers for CCAT (red points), shown with their respective $\pm3\sigma$ errors. The thick blue line is the best fit SPT foreground model, and the purple line is the lensed CMB power spectrum. The cyan and magenta shaded regions represent the $\pm3\sigma$ model uncertainties on the tSZ power spectrum from the SPT and CCAT, respectively. This figure illustrates how the improved sensitivity and angular resolution of CCAT can constrain both the amplitude and the shape of the tSZ power spectrum at the same time.}
\label{fig:ccat_bandpower}
\end{center}
\end{figure}

CCAT is expected to be a $25$ meter class submillimeter telescope that will perform high resolution microwave observations of the Southern sky \citep[e.g.,][]{Woo12}. It will enable accurate measurements of the tSZ and kSZ power spectra in the multipole range between $2000<\ell<20000$. CCAT will be more sensitive than SPT in the location of tSZ power spectrum peak, and thus can better constrain the shape and the normalization of the spectrum. Figure~\ref{fig:ccat_bandpower} shows simulated CCAT bandpowers at 150~GHz from a 5 years survey, performed over $2000~$deg$^{2}$ in approximately $10,000~$ hours of integration. The nominal noise value at 150 GHz for this fiducial CCAT survey is 12~$\mu$K/beam. It is assumed that the wide frequency coverage of CCAT, in particular its $850~\mu$m band, will effectively remove the dusty sub-mm galaxy confusion at lower frequencies. 

We have used predicted CCAT bandpowers created using the baseline SPT model (C. Reichardt, priv. comm.). Assuming the same shapes for the foreground power spectra templates, the models were extrapolated to higher $\ell$ values to account for the factor two better resolution of CCAT. For our analysis we also only used the three auto-spectra frequencies (95, 150 and 220 GHz) as in SPT, and the three cross-spectra, since the higher frequencies mostly provide better constraints on the CIB spectra. The survey area was scaled from the SPT survey area used in \citetalias{R12} for improved statistical errors. Calibration and the beam uncertainties were included at 5\% level. Although the increased frequency coverage of CCAT might enable a more precise modeling of the CIB background, we did not use any new foreground model for our predictions. The CCAT bandpowers thus reflect an experiment with better sensitivity and resolution but with our current knowledge of the microwave foreground templates.

As seen from Fig.~\ref{fig:ccat_bandpower}, the combination of unprecedented sensitivity and angular resolution of CCAT can constrain the shape and normalization of the tSZ power spectrum accurately, sufficient to break parameter degeneracies. When varying simultaneously the evolution parameter $\alpha_{z}$ and the slope term $\beta$, we obtain $\alpha_{z}=-1.42\pm0.07$ and $\beta=4.71\pm0.08$ (see Fig.~\ref{fig:predictions_SPTCCAT}, black contours). The marginalized errors on these two parameters thus show almost an order of magnitude improvement over the current SPT-based results. Similar tight constraints are obtained from other parameter combinations as well. This result is significant, since gaining an order of magnitude better accuracy through targeted observation of galaxy clusters, either in tSZ or in X-rays, will be very difficult, at least with the surveys planned for the coming decade. Through tSZ power spectrum measurements one can thus put the most stringent constraints on the mass and redshift scaling of the pressure profile in galaxy groups and clusters.
 
We can obtain very similar parameter constraints when using simulated bandpowers for the SPT-3G experiment. SPT-3G is the proposed third generation detector array on SPT \citep{Ben14}, and will possibly have marginally better sky sensitivity than CCAT due to its longer survey duration. However, its resolution will be worse than the CCAT and may not resolve the shape of the tSZ power equally well. It may also be less efficient in the modeling and removal of foreground components due to a smaller number of submillimeter frequency channels. Nevertheless, as we use the same frequency bands and the same baseline model templates for computing the CCAT and SPT-3G results, the respective model constraints turn out to be very similar. Our results here are not intended as a comparison between experiments, rather as a general demonstration of how these upcoming experiments can help to model cluster astrophysics parameters precisely through the tSZ power spectrum.

\begin{figure}
\begin{center}
\includegraphics[width=\columnwidth]{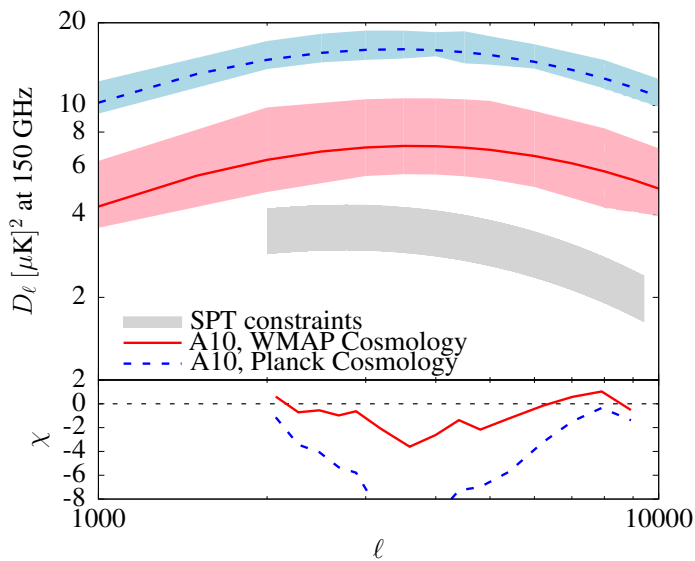}
\caption{Prediction for the tSZ power spectrum amplitude from the \citetalias{Ar10} model, but using both the WMAP7 and \plk best fit cosmological model parameters. The higher predicted amplitude from the \plk cosmology comes primarily from the higher values of $\sigma_8$ and $\Omega_\textrm{b}$. The shaded regions around the best-fit models are obtained using the respective parameters chains for these two parameters. The higher sensitivity of \plk clearly provides tighter constraints, although will require more drastic changes to the ICM pressure profile than we considered in this paper.}
\label{fig:cosmology}
\end{center}
\end{figure}

\subsection{Impact of cosmological uncertainties}
\label{sec:cosmology}

The key assumption in our work had been that cosmological parameters like $\sigma_8$ and $\Omega_{\textrm{m}}$ are known to infinite accuracies, which is not realistic. In this final Section we discuss the issue of parameter priors instead of fixed values. The error in cosmology can be of two different types. First, there is uncertainty in the cosmological model parameter fits in any given data set (or a combination thereof), that is given by the parameter covariances. Second, there can be additional systematic uncertainties between the best fit values from different probes, like that between the current WMAP and \plk results based on the CMB analysis. In Fig.~\ref{fig:cosmology} we show the difference in amplitudes of the tSZ power spectrum, computed using the \citetalias{Ar10} model without modifications, from either the WMAP7 \citep{Ko11} or \plk \citep{Pl15} best fit cosmological parameters. The roughly factor $2$ higher amplitude from \plk primarily comes from the higher values of $\sigma_8$ and $\Omega_\textrm{b}$, since the tSZ amplitude roughly scales as $D_{\ell}^{\mathrm{tSZ}} \propto \sigma_8^{8.3} \Omega_\textrm{b}^{2.8}$ \citepalias[e.g.,][]{R12}. Consequently, choosing the present \plk cosmological parameters instead of the WMAP values would require all the pressure profile modification results presented in this paper to be even stronger.

It is not the purpose of this work to address the current tension between the WMAP and \plk cosmological parameters values. However, even if a concordance is reached, there will always remain the statistical uncertainties (and some unresolved systematics) in any specific cosmological model that will affect the pressure model predictions based on the tSZ power. This issue can be addressed through applying known parameter priors in the MCMC chain while computing the halo mass function and the volume element.

We set priors on the two cosmological parameters that affect the tSZ power spectrum most: $\Omega_{\textrm{m}}$ and $\sigma_8$. The priors are from the WMAP9 measurements \citep{Hin13}, $\Omega_{\textrm{m}}=0.264\pm0.00973$ and $\sigma_8=0.81\pm0.014$, and we take care of the correlation between the parameters by using the actual parameter chains from WMAP9. We run our chains marginalizing over these two parameters, to constrain the redshift evolution parameter $\alpha_{z}$, as well as the CIB amplitude parameters. We obtain $\alpha_{z}=-0.98\pm0.25,~D^\textrm{P}_{3000}=7.69\pm0.24,~D^\textrm{C}_{3000}=6.36\pm0.48$, whose values and errors are consistent with the ones obtained previously (see subsection~\ref{subsec:nonSSz}). The use of \plk cosmological priors instead of WMAP9 provides a roughly factor 2 better constraints on these parameters, as can be seen from the respective uncertainty intervals in Fig.~\ref{fig:cosmology}. A similar conclusion was obtained by \citet{Hill13}, who obtained constraints for the outer-slope parameter $\beta$ at $6\%-8\%$ level, after marginalizing over cosmology, using a noise power model for \plk.

For the predicted CCAT bandpowers, we constrain the evolution parameter $\alpha_{z}$ in a similar way, with and without priors on the cosmological parameters. For the priors in this case we take a fiducial $1\%$ uncertainty on $\sigma_8$ and $\Omega_{\textrm{m}}$. This is assuming that by the time when CCAT will be in operation, the constraints on the cosmological parameters will be tighter thanks to some other experiments, like DES or eROSITA\footnote{http://www.mpe.mpg.de/eROSITA}. The results are displayed in Table~\ref{table:CCATpriors}. Clearly, switching from fixed cosmological values to realistic priors makes no major changes in the results, the same being true also for other pressure model parameters. The general conclusions presented in this work remain valid even when the parameter uncertainties are degraded by a factor $\sim 2-3$ while marginalizing over the cosmology.

\begin{table}
\caption[]{Comparison of the redshift evolution parameter $\alpha_z$ (Eq.~\ref{pe_alphaz}) and the CIB amplitudes with and without priors on the cosmological parameters. The adopted cosmology is from WMAP9 \citep{Hin13}, and we use the corresponding chains for cosmological parameters as priors, instead of fixed parameter values.}
\centering
{\small
\begin{tabular}{ c c c c c }
\hline
 & \multicolumn{2}{c}{SPT} & \multicolumn{2}{c}{CCAT} \\
\hline
 & No priors & With priors & No priors & With priors \\
\hline
$\alpha_{z}$ & $-0.73\pm0.16$ & $-0.98\pm0.25$ & $-0.73\pm0.02 $ & $-0.79\pm0.07$ \\
$D_{3000}^{\textrm{P}}$ & $7.69\pm0.27$ & $7.69\pm0.24$ & $7.69\pm0.01$ & $7.54\pm0.02$ \\
$D_{3000}^{\textrm{C}}$ & $6.35\pm0.49$ & $6.36\pm0.48$ & $6.35\pm0.04$ & $6.26\pm0.05$ \\
\hline
\end{tabular}}
\label{table:CCATpriors}
\end{table}

%%%%%%%%%%%%%%%%%%%%%%%%%%%%%%%%%%%%%%%%%%%%%%%%%%%%%%%%%%%%%%%%%%%%%%%%%%%%%%%%%%%%%%%%%%%%%%%%%%%%%%%%%%%%%%%%%%%%%%%%%%%%%%%%%%%%%%%%%%%%%%%

\section{Summary and conclusions}
\label{sec:summary}

We have provided a systematic calibration of intracluster gas models against observational data for the tSZ power spectrum. In particular, we used the GNFW model for an analytical description of the gas pressure profile with empirically determined parameters from \citet[][A10]{Ar10}. We tested various extensions of this model against the SPT measured values of CMB bandpowers on arcminute scales \citep[R12]{R12}. We employed an MCMC based method following the baseline model of SPT to explore the parameter likelihoods.

Similar to earlier works, we found that the ``universal'' pressure model of \citetalias{Ar10} produces an amplitude of the tSZ power spectrum that is roughly a factor two higher than that measured by the SPT, ACT, and {\it Planck}. In addition to the \citetalias{Ar10} model itself, we tested the GNFW models fitted directly to the \plk and {\it Bolocam} data, which fail to account for the low tSZ power in the same way as the \citetalias{Ar10} model. The measurement errors in the \plk and {\it Bolocam} results are small compared to the current mismatch between model predictions and experimental results.

We considered three different modifications to the \citetalias{Ar10} pressure model: first, following a strictly self-similar evolution; second, applying a weakly self-similar evolution where only the amplitude of the pressure profile changes with redshift and mass; and third, having a non-self-similar evolution where both the amplitude and shape of $P(r)$ change with redshift. For the self-similar case, we only varied the cluster outer slope parameter, $\beta$, because it has the weakest observational constraint. The maximum likelihood value, which needed to reconcile model predictions with the SPT bandpowers, is $\beta=6.3\pm0.2$. This is significantly higher than the most probable values measured by the {\it Bolocam} and \plk cluster tSZ observation. It also produces low-$\ell$ tSZ power that is inconsistent with the \plk tSZ bandpower measurements.

In a weak departure from self-similarity, we took the shape of the pressure profile as constant with redshift, but let the amplitude evolve differently than for self-similar models. We considered a power-law dependence of $(1+z)$ or an additional exponent to $E(z)$ to model this evolution in the pressure scaling. Such a dependence on redshift could be due to an evolution of the gas mass fraction, $f_{\gas}$, with redshift. We found that such models produce an excellent fit to the SPT data. However, an evolution of $f_{\gas}$ also affects the X-ray luminosity and would thereby produce some tension with the measured $L_X-T_X$ scaling relation of high-$z$ clusters. Additionally, a strong decrease in $f_{\gas}$ with redshift appears to be inconsistent with some recent hydrodynamical simulations of cosmological halo samples.

In a final attempt to modify the GNFW pressure profile of \citetalias{Ar10}, we let both its shape and amplitude vary with redshift in a strong departure from self-similarity. We considered a simple modeling for the redshift dependence of the pressure outer slope parameter $\beta$, as motivated by the recent simulations of random gas motions in the cluster outskirts. We found that in such cases the parameters are highly degenerate: neither the pressure slope at $z=0$ nor its redshift evolution can be constrained accurately from the current tSZ power spectrum data.

The degeneracy between the model parameters is a general problem when using the currently available CMB bandpower measurements. For future CMB/tSZ experiments with better sensitivities, these degeneracies can be broken by measuring both the shape and the amplitude of the tSZ power spectrum to high accuracy. We used the simulated bandpower measurements for a CCAT 2000 deg$^{2}$ sky survey and found almost an order~of~magnitude improvement over the current model parameter uncertainties. This can, for example, enable simultaneous measurements of the outer slope parameter and the redshift evolution of $f_{\gas}$ to the level of a few percent.

We tested the impact of cosmological parameter uncertainties, in particular $\sigma_8$ and $\Omega_\textrm{m}$, on our results. For the current SPT-data based results, we used the WMAP9 cosmological parameter uncertainties, directly using the chains for the relevant cosmological parameters as priors in our MCMC modeling. This degrades the uncertainties on the pressure profile parameters like $\beta$ or $\alpha_{z}$ by roughly a factor two. For the CCAT fiducial survey, we reduced the cosmological errors by an additional 50\%. This causes the ICM pressure model errors for CCAT to degrade roughly by a factor three, which will still be sufficient to place strong constraints on multiple model parameters. The large systematic difference between the current WMAP and \plk cosmological parameters remain an open question, although we note that adopting to the \plk cosmology will roughly cause a factor two higher prediction of the tSZ power amplitude, so will require more drastic modifications to the ICM pressure profile than we considered in this paper.

%%%%%%%%%%%%%%%%%%%%%%%%%%%%%%%%%%%%%%%%%%%%%%%%%%%%%%%%%%%%%%%%%%%%%%%%%%%%%%%%%%%%%%%%%%%%%%%%%%%%%%%%%%%%%%%%%%%%%%%%%%%%%%%%%%%%%%%%%%%%%

\section*{Acknowledgements}
We acknowledge the help from Christian Reichardt and Daisuke Nagai in providing the simulated bandpower measurements for CCAT; Laurie Shaw for sharing his tSZ and kSZ power spectrum templates; and Jack Sayers for sharing the {\it Bolocam} cluster pressure profile fit results. The authors acknowledge financial support from the DFG (Deutsche Forschungsgemeinschaft), through the Transregio Programme TR33 ``The Dark Universe''. FP acknowledges support from BMBF/DLR grant 50~OR~1117.

{\small
\bibliographystyle{aa}
\bibliography{tSZps_references}
}

%%%%%%%%%%%%%%%%%%%%%%%%%%%%%%%%%%%%%%%%%%%%%%%%%%%%%%%%%%%%%%%%%%%%%%%%%%%%%%%%%%%%%%%%%%%%%%%%%%%%%%%%%%%%%%%%%%%%%%%%%%%%%%%%%%%%%%%%%%%%%

\appendix

\section{Details of the R12 sky model}
\label{app:A}

In this appendix, we summarize the models defined by \citet[R12]{R12} for the different components of the microwave sky diffuse emission. These are used as a baseline throughout the article while constraining parameters from the measured SPT bandpowers.

\begin{itemize}
\item {\it Lensed Primary CMB}. The lensed CMB is calculated with CAMB\footnote{http://lambda.gsfc.nasa.gov/toolbox/tb\_camb\_form.cfm}. Gravitational lensing effects are important because they tend to increase the power of the CMB anisotropies on small angular scales, compared to the unlensed estimates.

\item {\it Poisson infrared (IR) source power}. This term takes into account the shot-noise fluctuation power from randomly distributed microwave-emitting galaxies. It is given by 
\begin{equation}
D_{\ell,\nu_{1},\nu_{2}}^{\textrm{P}}=D_{3000}^{\textrm{P}}\epsilon_{\nu_{1}\nu_{2}}\eta_{\nu_{1}\nu_{2}}^{\alpha_{\nu}}\bigg(\frac{\ell}{3000}\bigg)^2,
\end{equation}
where $D_{3000}^{\textrm{P}} \equiv D_{3000,\nu_{0},\nu_{0}}^{\textrm{P}}$ is the amplitude of the Poisson power term for infrared galaxies at $\ell=3000$ and at reference frequencies ($\nu_0$): $97.9$, $153.8$, and $219.6$ GHz. The coefficient
\begin{equation}
\epsilon_{\nu_{1},\nu_{2}}\equiv \frac{\frac{dB}{dT}|_{\nu_{0}}\frac{dB}{dT}|_{\nu_{0}}}{\frac{dB}{dT}|_{\nu_{1}}\frac{dB}{dT}|_{\nu_{2}}},
\end{equation}
is the ratio of power in the $\nu_{1}\otimes \nu_{2}$ cross-spectrum with respect to the $\nu_{0}\otimes \nu_{0}$ auto-spectrum. $\eta_{\nu_{1}\nu_{2}}=(\nu_{1}\nu_{2}/\nu_{0}^{2})$ is the ratio of the frequencies of the spectrum to the reference frequency. \citetalias{R12} obtained a best fit value for the spectral index of $\alpha_{\nu}=3.45$. $B(T)$ is the CMB blackbody specific intensity.

\item {\it Clustered IR source power}. Because the infrared galaxies trace the mass distribution, they are spatially correlated. This leads to a clustered term in the power spectrum of infrared galaxies given by
\begin{equation}
D_{\ell,\nu_{1},\nu_{2}}^{\textrm{C}}=D_{3000}^{\textrm{C}}\epsilon_{\nu_{1}\nu_{2}}\eta_{\nu_{1}\nu_{2}}^{\alpha_{c}}\bigg(\frac{\ell}{3000}\bigg)^{0.8},
\end{equation}
where $D_{3000}^{\textrm{C}}$ is defined as in the IR Poisson term, and $\alpha_{c}=3.72$ is the best fit value taken from \citetalias{R12}. Moreover, \citetalias{R12} adopted the power law model $\ell^{0.8}$ from Lyman-break correlated galaxies \citep{Gia98,SW99}.

\item {\it Radio source power}. The brightest point sources in the SPT maps coincide with known radio sources. To take this contribution into account, the Poisson radio term is given by
\begin{equation}
D_{\ell}^{\textrm{R}}=D_{3000}^{\textrm{R}}\epsilon_{\nu_{1}\nu_{2}}\eta_{\nu_{1}\nu_{2}}^{\alpha_{\textrm{R}}}\bigg(\frac{\ell}{3000}\bigg)^2,
\end{equation}
where $D_{3000}^{\textrm{R}}$ is the amplitude of the radio Poisson power spectrum at $\ell=3000$ with value $1.28\pm0.19~\mu$K$^2$ at $150$~GHz. This value is based on the \citet{Zo05} source count model. $\alpha_{\textrm{R}}=-0.53$ is the mean spectral index from the \citet{Sh11} analysis.

\item {\it Galactic cirrus}. An average Galactic cirrus contribution term is parametrized as 
\begin{equation}
D_{\ell}^{\textrm{cir}}=D_{3000}^{\textrm{cir},\nu_{1},\nu_{2}}\bigg(\frac{\ell}{3000}\bigg)^{-1.2}.
\end{equation}
Following the cirrus treatment in \citet{Ha10} and \citet{Ke11}, \citetalias{R12} measured powers at $\ell=3000$ to be $0.16$, $0.21$, and $2.19~\mu$K$^2$ for $95$, $150$, and $220$~GHz respectively.

\item {\it kSZ power spectrum}. A homogeneous kSZ power spectrum is adopted, following the {\it cooling plus star-formation} (CSF) model of \citet{Sh12}. This model is constructed by calibrating an analytic model with two hydrodynamical simulations. 
The CSF template predicts $1.6~\mu$K$^2$ at $\ell=3000$. 

\item {\it tSZ power spectrum}. \citetalias{R12} uses the analytical model of \citet{Sh10} as a template for the tSZ power, which relies on a physical cluster model coupled with halo formalism similar to the one presented in Section~\ref{subsec:SZcompute}. Their cluster model mainly accounts for star formation, energy feedback (from supernovae and active galactic nuclei), and non-thermal pressure support. In our analysis, we replace the Shaw et al. model by a phenomenological description of the intracluster pressure profile that allow us for more freedom.
\end{itemize}

\label{lastpage}

\end{document}